\documentclass[11pt,a4paper]{article}

\usepackage{amsmath} 
\usepackage{amssymb} 
\usepackage{makeidx} 
\usepackage{graphicx} 
\usepackage{pdflscape}
\usepackage{rotating}

\usepackage{subfigure}
\usepackage{geometry}
\usepackage{caption}
\usepackage{hyperref}

\begin{document}

\captionsetup{width=0.85\textwidth}

\author{A. Reiter, R. Singh, O. Chorniy \\ Beam Instrumentation Department \\ GSI Helmholtz Centre for Heavy Ion Research}
\title{ Statistical Treatment \\of Beam Position Monitor Data}
\date{\today}
\maketitle

\abstract
We review beam position monitors adopting the perspective of an analogue-to-digital converter in a sampling data acquisition system.\\ 

From a statistical treatment of independent data samples we derive basic formulae of position uncertainty for beam position monitors. Uncertainty estimates only rely on a few simple model parameters and have been calculated for two "practical" signal shapes, a square pulse and a triangular pulse. The analysis has been carried out for three approaches: the established signal integration and root-sum-square approaches, and a least-square fit for the models of direct proportion and straight-line. The latter approach has not been reported in the literature so far.\\

The three approaches lead to identical position estimates, if no noise contribution and proper baseline restoration are assumed in the integration approach. A significant advantage of the fit approach is the fact that baseline samples can be included in the calculation without adverse effects while they increase the uncertainty in the integration method. More importantly, the fit approach eliminates the need for baseline restoration which greatly simplifies the data handling. The RSS approach turns out to be equivalent to a direct proportion fit and, hence, also does not require baseline restoration. But, like the integration approach, it suffers from external sources of signal distortions which are dominant at low frequencies and can lead to systematic effects that are difficult to detect and quantify.\\

The straight-line fit provides the most robust estimator since it does not require baseline restoration, it is immune to signal offsets, and its standard deviation is smallest. Consequently, of the analysed estimators it promises the highest fidelity of results. The fit approach represents a simple, natural way to analyse beam position monitor data which can be easily implemented in hardware and offers potential for new applications.

\newpage
\tableofcontents
\newpage

\section{Position Monitor Data Analysis: A dead horse?}

During our search in the wealth of literature about beam position monitors (BPM) we encountered numerous documents that cover seemingly all aspects of this type of detector (see~\cite{1,2,3,4,5,6,7} and references therein): rf characteristics of different geometries and mechanical assemblies, estimates of detector sensitivity and response to the passing ensemble of charged particles, or prescriptions how to calculate the mean beam position. All questions seemed to have been dealt with, until we tripped over the basic topic of measurement uncertainty, i.e. the question of how well we determine the beam position in reality and of how to interpret this contribution in position measurements in the daily operation of a heavy-ion synchrotron. \\
Many publications treat the BPM problem, but we were missing a simple treatment which allows to quantify the quality of the acquired data set, i.e. to assign an uncertainty to a single position  measurement. After all, BPM signals are nowadays often acquired in ADC sampling systems and the beam position is evaluated from these raw data. Therefore, the most basic unit of information is a single ADC datum or sample with a given noise level or uncertainty.\\
For this reason we embark in this monograph on a "statistical journey" of BPM data analysis. We want to evaluate and compare the well known prescriptions on position measurement and we hope to find an answer to the following question:\\
{\it \bf Which is our best position estimator judged by robustness and uncertainty of the result?}\\

To this end we have calculated beam positions from the most common algorithms in a simple theoretical model. Although some results seemed obvious, finally the statistical treatment was very instructive and gave insight into the strengths and weaknesses of different estimators and into their relationship. Most importantly, the examination of this model, together with the results obtained from simulated and real data, convinced us to introduce a new analysis approach to all further applications: the asynchronous mode. Aside, the theoretical model produced the practical uncertainty estimates that had been initially our primary objective. So we believe that we were not flogging a dead horse after all! \\

In section~\ref{SecElectronics} a typical ADC system for the acquisition of BPM signals is described. Section~\ref{SecBeamPosition} introduces the fundamental "difference-over-sum" equation for position evaluation. To this equation we apply all theoretical beam models and algorithms presented in section~\ref{SecCalcPosition} and discuss the characteristics of the estimators. We proceed in section~\ref{SecCalcUncertainty} with the questions that had sparked our effort, i.e. the calculation of the position uncertainty. At the end of this sections all results are compared and some conclusions are drawn for future implementations of BPM measurement systems.

\section{BPM Electronics}
\label{SecElectronics}

\begin{figure}[h]
\begin{center}
\includegraphics[width = 100mm]{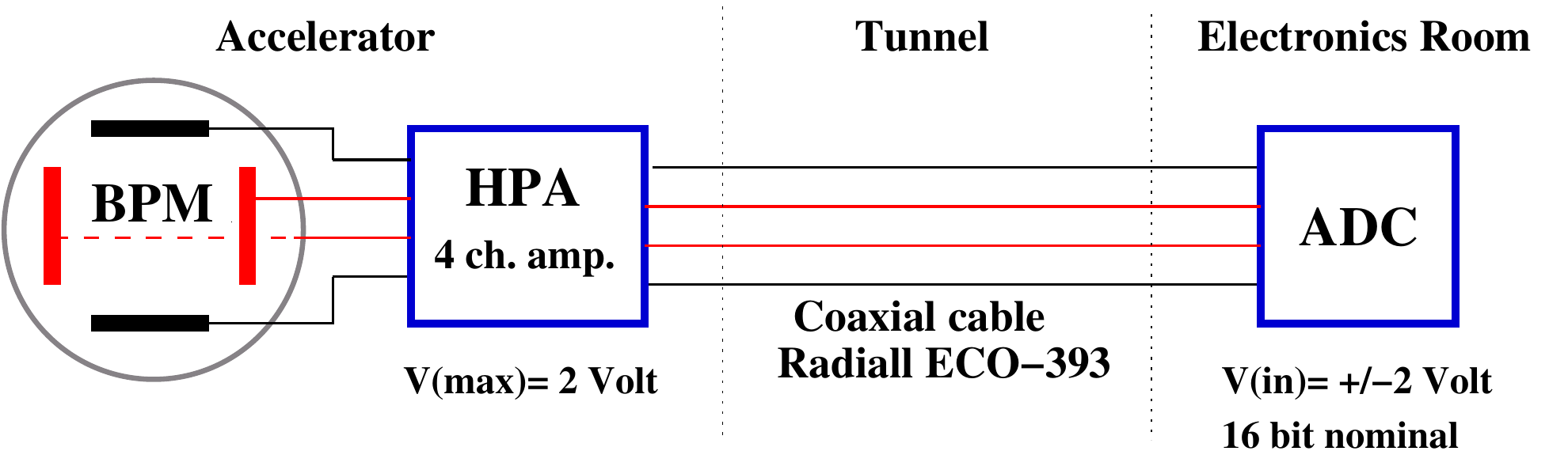}
\caption{Schematic of BPM data acquisition system consisting of BPM with horizontal and vertical electrode pairs, four-channel head preamplifier HPA and ADC sampling system.}
\label{BPM_Schematics}
\end{center}
\end{figure}

We consider a diagonally-cut cylindrical BPM that measures the beam position in the horizontal plane. For example calculations we use the parameters of a BPM with 100 mm diameter (radius $r$ = 50 mm), the detector type installed in the experimental storage ring CRYRING@ESR. Figure~\ref{BPM_Schematics} illustrates the hardware setup consisting of BPM electrode pair, preamplifier, coaxial transmission line and ADC system.\\
The two BPM electrodes are supposed to be connected to a matched amplifier pair, i.e. amplifiers of identical gains, and an ideal ADC of fixed input range and maximum voltage $V_{FS}$. Left and right electrode signal are called $S_L(t)$ and $S_R(t)$, respectively, and are functions of the ADC sample number $i$ or sample time $t=i\cdot t_{Sa}$ where $t_{Sa}$ is the sampling interval. Note that typically the BPM and the following amplifier form an AC coupled system whose lower cut-off frequency is given by the input impedance of the amplifier. The AC coupling leads to baseline shifts or offsets for repetitive signals if the repetition time is much lower compared to the inverse of lower-cut off frequency.\\
The RMS noise voltage $\sigma_V$, that is the uncertainty of a single ADC datum, is defined as the standard deviation of a baseline (or offset) measurement performed when there is no external stimulus at the BPM electrodes. Hence, this definition of the uncertainty $\sigma_V$ includes all noise contributions along the signal chain. We assume a constant value for $\sigma_V$, independent of the measured signal level. However, we should note that the noise characteristics of a realistic amplifier and ADC cannot be fully specified by a single number $\sigma_V$ since the frequency spectra are not "white" and therefore $\sigma_V$ is not the same for all position analysis methods. The methods looking into lower frequency region will suffer from higher noise compared to the ones looking into higher frequency regimes. More details on this aspect is reserved for another report.

\section{Beam Position \& Relative Signal Levels}
\label{SecBeamPosition}
For the diagonally-cut cylindrical BPM the mean beam position $x$ can be derived from the two signals $S_L$ and $S_R$ of left and right pickup electrode and the known BPM radius r: 
\begin{equation}\label{EquTheory}
\boxed{x(t) / r = \frac{\Delta(t)}{\Sigma(t)} = \frac{S_L(t) - S_R(t)}{S_L(t) + S_R(t)}}
\end{equation}
For more information refer to references~\cite{1,2,4}.
We rewrite this formula to obtain a relation between $S_L$ and $S_R$ as function of the beam position $x$. We define the normalised position $\alpha=x/r$, the beam position or offset in units of the BPM radius defined in the range [-1,+1], and derive:
\begin{equation}
\boxed{S_L =\big( \frac{1+\alpha}{1-\alpha}\big) \cdot S_R = c(x)\cdot S_R} \label{EquSLSR}
\end{equation}
We will frequently make use of equation~\ref{EquSLSR}, i.e. of the direct proportion between left and right electrode signal, throughout this document. A few examples are given below

\begin{eqnarray}
\textrm{For small $\alpha$:\quad}S_L&\sim& (1+2\alpha+ 2\alpha^2)\cdot S_R\\
\alpha = 0.005:\quad S_L &=& 1.01 \cdot S_R \textrm{\quad(0.09 dB)}\nonumber\\
\alpha = 0.01:\quad S_L &=& 1.02\cdot S_R\textrm{\quad(0.17 dB)}\nonumber\\
\alpha = 0.05:\quad S_L &=& 1.11\cdot S_R\textrm{\quad(0.86 dB)}\nonumber\\
\alpha = 0.1:\quad S_L &=& 1.22\cdot S_R\textrm{\quad(1.72 dB)}\nonumber\\
\alpha = 0.2:\quad S_L &=& 1.50\cdot S_R\textrm{\quad(3.50 dB)}\nonumber
\end{eqnarray}
{\it Therefore beam position measurements need to detect small differences of the order of a few percent between two large signals, since the beam is held close to the reference orbit (\,$\lvert x/r \rvert <<1$\,) in accelerators and transport beam lines. This poses strict requirements on the quality of amplifier and ADC hardware in order to achieve proper matching and good noise characteristics of the electronics.}

\section{Calculation of Beam Position}
\label{SecCalcPosition}
We discuss three approaches for position evaluation: the common approaches of signal integration and root-sum-square RSS, and a new approach that evaluates the position from a least-square fit to the two signals. Trial functions are the direct proportion and the straight-line. For the signal integration it is assumed that both signal traces $S_L(t)$ and $S_R(t)$ are free of baseline drifts or that their baseline has been "perfectly" restored.\\
We will calculate the position for two pulse shapes, a rectangular and a triangular pulse, because these pulses can be handled analytically without great effort. Further, the triangular pulse shape seems a good approximation for many practical cases. More complex beam pulses, e.g. a circulating beam or a train of 2, 4 or 8 pulses extracted from a synchrotron, can be constructed on the basis of the single-bunch model.  

\subsection{Beam Models}
\label{SecBeam}
\subsubsection{Square pulse without baseline offset}
\label{SecSquare}
\begin{figure}[h]
\begin{center}
\includegraphics[width = 0.75\textwidth]{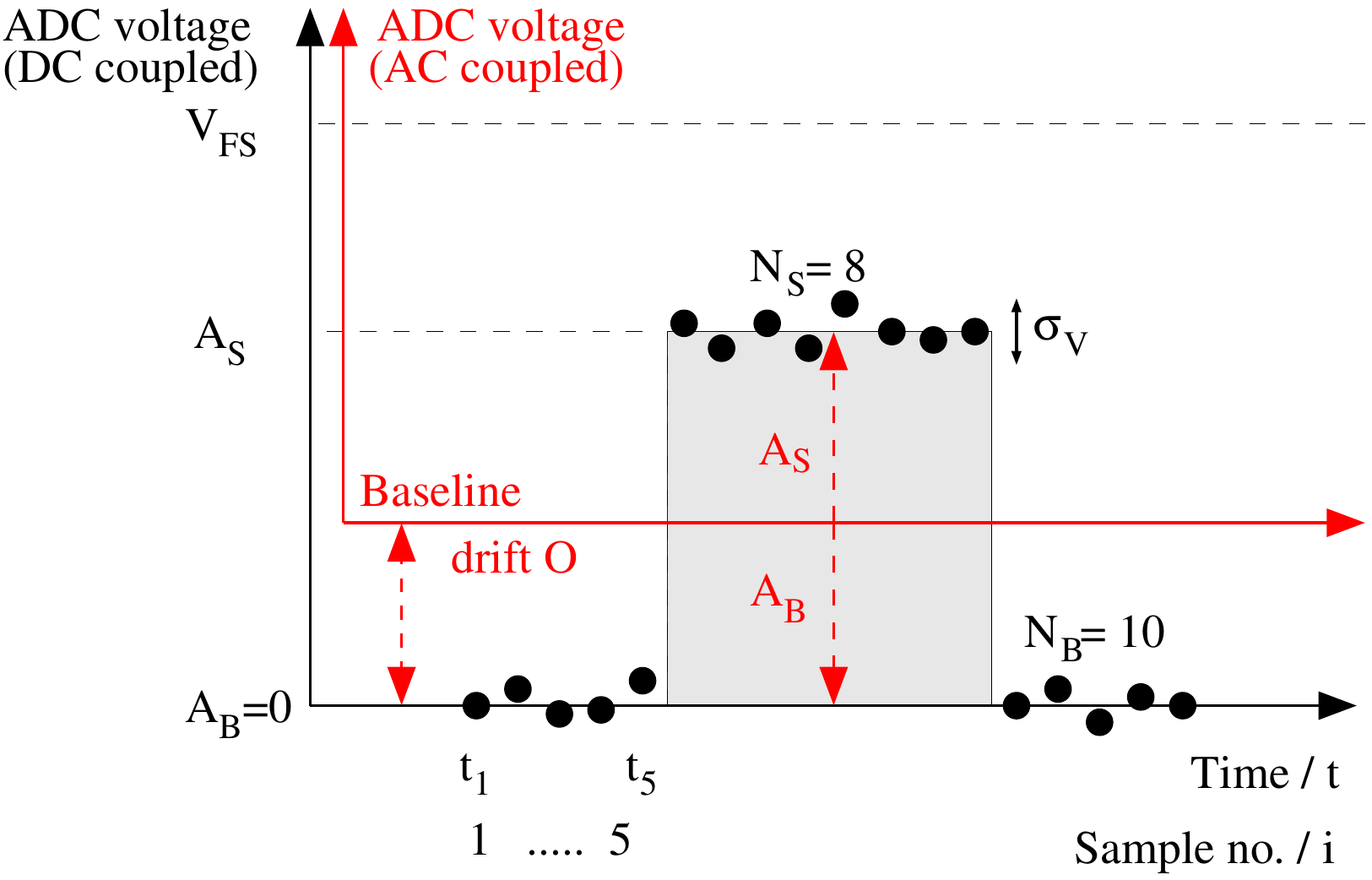}
\caption{ADC data composed of baseline ($N_B=10$) and a single square pulse ($N_S$=8). The black coordinate system corresponds to a DC coupled system (or a single-pass measurement), while the red one represents the AC coupled case. In a "multi-pass" measurement, the baseline initially drops and finally stabilizes at offset $O$.}
\label{FigSquare}
\end{center}
\end{figure}

The rectangular pulse is illustrated in Fig.~\ref{FigSquare} and the black coordinate system refers to the case without baseline offset. The acquisition window is larger than the pulse width and includes some baseline samples. We call the number of "signal samples" that carries information on the signal $N_S$ and the number of "baseline samples" $N_B$. The maximum signal amplitude is expressed by a dimensionless parameter A in units of the full scale voltage $V_{FS}$. For the i-th signal sample the amplitude $S(i)$ is simply: $S(i) = A_S\cdot V_{FS}$.

\subsubsection{Square pulse with baseline offset}
\label{SecSquareAC}
The AC coupling in the electronics chain generates a baseline shift or offset $O$ because the DC blocking characteristics forces the areas above and below the zero-line to be equal. In Fig.~\ref{FigSquare} this effect  is illustrated by the red coordinate system.
Assuming a constant offset $O$ between two successive pulses of length $t_s$ which are separated by time period $t_p$,  one can calculate the baseline offset for the square pulse as:
\begin{eqnarray}
\int_{t_0=0}^{t_p} (S(t)-O)\, dt &=& \int_{t_0=0}^{t_s} S(t)\,dt   - \int_{t_0=0}^{t_p} O\, dt = I - O\, t_p = 0
\nonumber
\end {eqnarray}
\begin{equation}
\boxed{O = \frac{I} { t_p}}
\label{EquOffset}
\end{equation}

In sample space there are $N_S$ signal samples followed by $N_B$ baseline samples until the next pulse arrives. $N_S+N_B$ represents the total number of samples between two successive pulses and the repetition period $t_p=(N_S+N_B)\cdot t_{Sa}$. Then the offset O is calculated as:
\begin{eqnarray}
  \sum_{i=1}^{N_S+N_B} (S_i-O) &=& \sum_{i=1}^{N_S} S_i- \sum_{i=1}^{N_S+N_B} O = I - \sum_{i=1}^{N_S+N_B} O = 0 \nonumber \\
   \sum_{i=1}^{N_S+N_B} O &=&  \sum_{i=1}^{N_S} S_i = I \label{EquOffCalc}
\end{eqnarray}

\begin{equation}
\boxed{O =    \frac{I}{N_S+N_B}= \frac{ N_S \cdot (A\cdot V_{FS})}{N_S+N_B}}\label{EquOffSqu}
\end{equation}

\subsubsection{Triangular pulse without baseline offset}
\label{SecTriangle}
\begin{figure}[h]
\begin{center}
\includegraphics[width = 100mm]{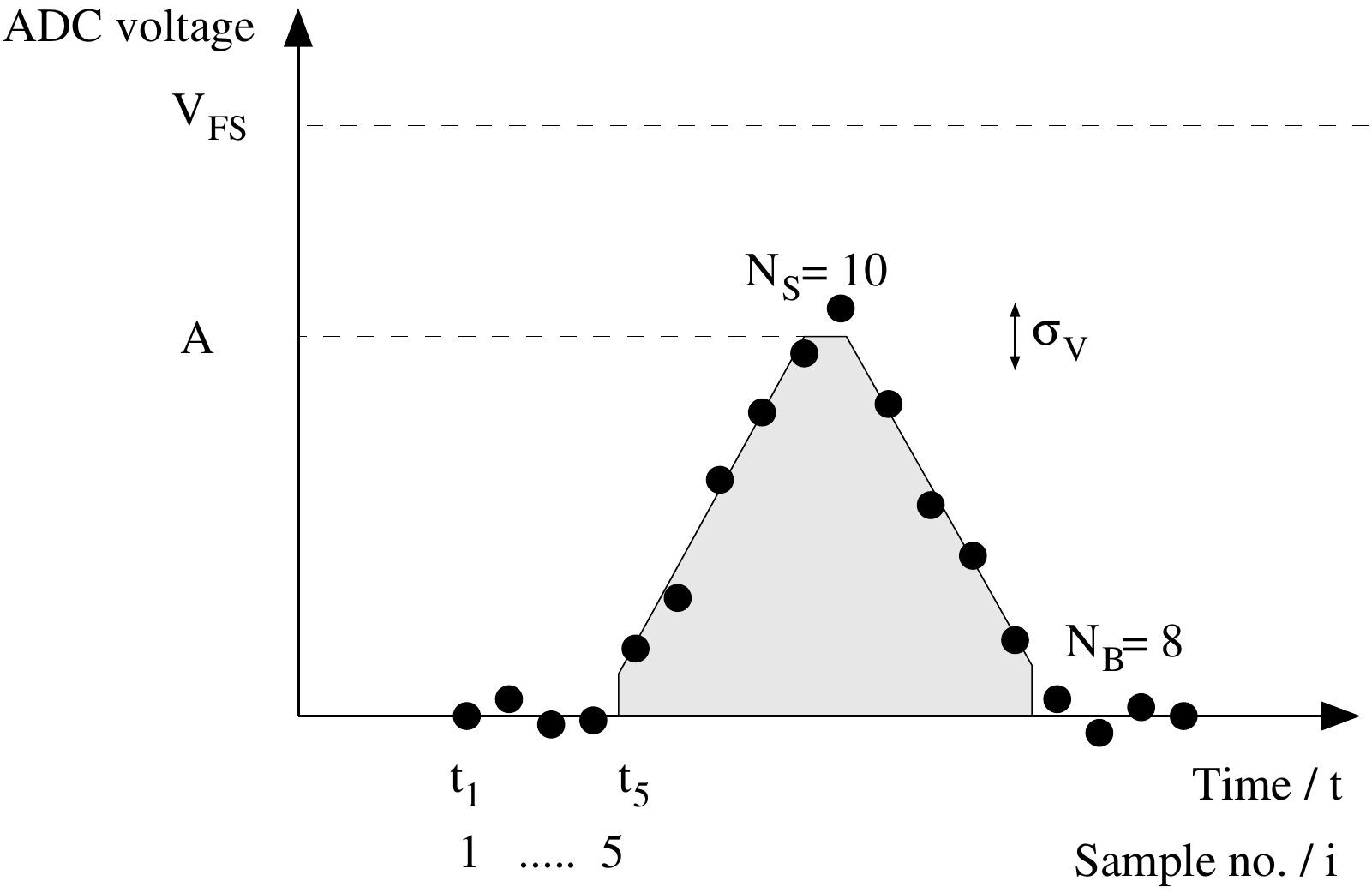}
\caption{ADC data composed of baseline ($N_B=8$) and a single triangular pulse ($N_S$=10). Note that we assume two, rather than one, samples at the peak to guarantee an even sample number.}
\label{FigTriangle}
\end{center}
\end{figure}

The triangular pulse is illustrated in Fig.~\ref{FigTriangle}. The acquisition window is larger than the pulse width and includes some baseline samples. For the sake of simplicity in the analytical calculation two data samples are assumed at the peak to guarantee an even sample number, and identical rising and falling edges.\\
For the i-th signal sample the amplitude S(i) is given by: 
\begin{eqnarray}
S(i) &=&  i\cdot \frac{2\cdot A \cdot V_{FS}}{N_S} \quad\quad\quad\quad\quad\quad\textrm{for i $\in$ [1, $N_S$/2]} \\
S(i) &=&  (N_S+1-i) \cdot \frac{2\cdot A \cdot V_{FS}}{N_S} \quad\textrm{for i $\in$ [$N_S$/2+1, $N_S$]}
\end{eqnarray}
The signal integral $I$ is given by the triangle area (see equation~\ref{EquTriInt} of section~\ref{SecUncertaintyInt}):
\begin{equation}
I = \frac{N_S+2}{2} \cdot (A\cdot V_{FS}) \nonumber
\end{equation}

\subsubsection{Triangular pulse with baseline offset}
\label{SecTriangleAC}
Assuming a constant offset $O$ between two successive pulses of length $t_s$ which are separated by time period $t_p$, one can calculate the baseline droop for the triangular pulse again from equation~\ref{EquOffset}:
\begin{equation}
O = \frac{I} { t_p}\nonumber
\end{equation}

In sample space there are $N_S$ signal samples followed by $N_B$ baseline samples until the next pulse arrives. $N_S+N_B$ represents the total number of samples between two successive pulses and the pulse repetition period $t_p=(N_S+N_B)\cdot t_{Sa}$. Then the offset O is calculated:
\begin{eqnarray}
   \sum_{i=1}^{N_S+N_B} O &=&  \sum_{i=1}^{N_S} S_i = I \label{EquDroop1}
\end{eqnarray}

\begin{equation}
\boxed{O =    \frac{I}{N_S+N_B}= \frac{( N_S+2 ) \cdot (A\cdot V_{FS})}{2\cdot (N_S+N_B)}}\label{EquOffTri}
\end{equation}\\

\subsection{Classical Approach 1: Signal Integration}
\label{SecIntegration}
\subsubsection{Position Calculation}
In practical applications, one is interested in the position estimate $<x>$ of the complete bunch and integrates over the time-dependent electrode signals separately. Equation~\ref{EquTheory} is therefore modified to yield a single value:
\begin{equation}\label{Equ1}
\frac{<x>}{r} =  \frac{\int\Delta(t) \,dt}{\int\Sigma(t) \,dt} = \frac{I_L - I_R}{I_L + I_R}
\end{equation}
where the integrals $I_L$ and $I_R$ are calculated from the ADC data of the signals traces $S_L(t)$ and $S_R(t)$ . We stress again, that a proper offset correction or baseline restoration is absolutely vital to avoid biased results as discussed in the next section.

\subsubsection{Baseline restoration}
\label{BaselineRestoration}
The effect of baseline shift is caused by the AC coupling of the measurement system. The need for baseline restoration depends on the application: it may not be necessary if a short train of several bunches travels along a transfer beam line, but it may be absolutely crucial for synchrotron BPMs as discussed in section~\ref{SecElectronics}. In the latter application the baseline shift depends on the signal dynamic during the synchrotron cycle. It should be mentioned that, since DC is completely lost in an AC coupled system by definition, a perfect restoration of the DC or baseline is not possible using a linear operation. \\ 
Baseline restoration introduces a correlation between the offset-corrected data samples of a given signal, and increases the uncertainties of the corrected signals $S_L$ and $S_R$ (see section~\ref{SecUncertaintyBLR}). Further, and more importantly, systematic effects may be introduced that are very difficult to quantify in the practical application. We now look at three simple examples:\\

{\it 1) Common signal offset:} Let us assume a slightly imperfect baseline restoration and a common offset $O$ in both signals. Then, a bias is introduced to the position calculation:
\begin{equation}
x = r \,\cdot (\Delta/\Sigma) = r \,\cdot \frac{S_L - S_R}{S_L + S_R + 2 \cdot O}
\end{equation}
Assuming a true 0.5 mm beam shift towards the left electrode, we get $S_L = 1.0202\cdot S_R$ from equation~\ref{Equ1}.  An offset of 1 \% with respect to the signal level, leads to a position estimate that is reduced by the same amount: 
\begin{equation}
x = r \,\cdot (\Delta/\Sigma) = r \,\cdot \frac{0.0202}{2.0202 + 2\,\cdot 0.01} = 0.495 \, \mathrm{mm}
\end{equation}

{\it 2) Asymmetric signal offset:} Let us now assume an offset $O$ in signal $S_L$ alone:
\begin{equation}
x = r \,\cdot (\Delta/\Sigma) = r \,\cdot \frac{S_L - S_R + O}{S_L + S_R + O}
\end{equation}
Assuming the same position offset of 0.5 mm as above, the following position estimate is calculated:
\begin{equation}
x = r \,\cdot (\Delta/\Sigma) = r \,\cdot \frac{0.0202 + 0.01}{2.0202+ 0.01} = 0.74 \, \mathrm{mm}
\end{equation}

The result deviates by 50\% from the true position, but this should not come as an surprise to us. An asymmetry measurement that analyses a small difference with respect to a large total signal is very sensitive to systematic effects. This situation is very typical for a BPM measurement where the beam is close to its reference orbit. \\

{\it 3) Slow movement of the beam:} If the beam has moved from 0-0.5 mm in the last 100 $\mu s$ (for a 10 kHz cut-off of capacitive pick-up with high impedance termination), which is typical during an acceleration cycle, the baseline will reflect an "averaged" position of 0.25 mm, i.e. $S_{Lb} = 1.0101\cdot S_{Rb}$.
If we calculate the position using baseline restoration,
\begin{equation}
x = r \,\cdot (\Delta/\Sigma) = r \,\cdot \frac{W_{s}(S_L - S_R) + W_{b}(S_{Lb}-S_{Rb})}{W_{s}(S_L + S_R) + W_{b}(S_{Lb}+S_{Rb})}
\end{equation}
where $W_{s}$ and $W_{b}$ are the relative weights given to signal samples and baseline samples.
\begin{equation}
x = r \,\cdot (\Delta/\Sigma) = r \,\cdot \frac{0.0202+0.0101}{2.0202 + 2.0101} = 0.35 \, \mathrm{mm}
\end{equation}

{\it We will see later that here lies the benefit of the new approach in producing unbiased and more stable position estimates, especially at low beam intensities, compared to the classical approaches.}

\subsection{Classical Approach 2: RSS Calculation}
\label{SecRSS}
\subsubsection{Position Calculation}

In this approach, the position estimate $<x>$ is obtained by calculating the "root-sum-square" values, $RSS_L$ and $RSS_R$, of each time-dependent electrode signal $S_L(t)$ and $S_R(t)$ separately:

\begin{equation}\label{eq:pow}
\frac{<x>}{r} = \frac{RSS_L - RSS_R}{RSS_L + RSS_R}
\end{equation}


where the $RSS$ value of a signal is calculated from the ADC samples $S(i)$:

$$ RSS_{L/R}=\sqrt{\Sigma_{i=1}^N{S_{L/R}(i)}^2} $$
 
This method does not require the restoration of the DC baseline of the signal, which is lost due to AC coupling. However, uncontrolled offsets or drifts on either electrode lead to systematic biases. 

\subsubsection{Equivalence to Signal Integration}\label{SecRss=Int}
It might not be obvious that the two classical approaches yield the same position estimate. Using the direct proportionality $S_L= c(x)\cdot S_R$ for a fixed position $x$ given by equation~\ref{EquSLSR} the equivalence can be understood from a simplistic approach:

\begin{eqnarray}
\nonumber\left(\frac{<x>}{r}\right)_{Int} &=& \frac{I_L - I_R}{I_L + I_R} = \frac{\sum_i (S_L(i) - S_R(i))}{\sum_i (S_L(i) + S_R(i))}  \\ \nonumber
&=& \frac{\sum_i(c\cdot S_R(i) - S_R(i))}{\sum_i(c \cdot S_R(i) + S_R(i))} = \frac{(c - 1)\sum_i S_R(i)}{(c + 1)\sum_i S_R(i)} = \\ 
&=& \frac{c-1}{c+1} \\ \nonumber \\ \nonumber
\left(\frac{<x>}{r}\right)_{RSS} &=& \frac{RSS_L - RSS_R}{RSS_L + RSS_R} \\ \nonumber &=&\frac{\sqrt{\Sigma_{i=1}^N{S_L(i)}^2} - \sqrt{\Sigma_{i=1}^N{S_R(i)}^2}}{\sqrt{\Sigma_{i=1}^N{S_L(i)}^2} + \sqrt{\Sigma_{i=1}^N{S_R(i)}^2}}\\ \nonumber
&=& \frac{\sqrt{\Sigma_{i=1}^N{(c\cdot S_R(i))}^2} - \sqrt{\Sigma_{i=1}^N{S_R(i)}^2}}{\sqrt{\Sigma_{i=1}^N{(c\cdot S_R(i))}^2} + \sqrt{\Sigma_{i=1}^N{S_R(i)}^2}} = \\ 
&=&\frac{c-1}{c+1} 
\end{eqnarray}

\subsubsection{Effect of AC Coupling on RSS signal}
\label{SecRssAcSignal}

We re-calculate the RSS signal for a cyclic, triangular pulse train with baseline offset $O$ given be equation~\ref{EquOffTri}. We include in this analysis all samples within one period, $N=N_S+N_B$. This mimics the case of a turn-by-turn analysis  (harmonic number $h=1$) of a balanced BPM system where the baseline shift has settled or changes slowly compared to the revolution time:
\begin{align}  \label{EquAC-RSS}
\sum_{i=1}^{N} (S(i) - O)^2 &= \sum S(i)^2 - 2\cdot O \sum S(i) + \sum O^2 \nonumber \\ 
&= \sum S(i)^2 - 2\cdot O \cdot ( O \cdot N) + N \cdot O^2 \\ 
&= \sum S(i)^2 - N \cdot O^2 = \sum ( S(i)^2 - O^2) 
\end{align}

Note that we have used equation~\ref{EquDroop1} which defines the boundary condition for an AC coupled signal: the mean value is zero. Since offset $O$ is proportional to the integral signal strength, we expect that the value of the RSS signal is reduced by a factor which depends on pulse height and pulse separation. We apply equation~\ref{EquAC-RSS} to the triangular model without baseline offset and anticipate equation~\ref{EquRSS2} of section~\ref{SecUncRssTri}.

\begin{equation}
RSS^2 = \sum_{i=1}^{N_S+N_B} S(i)^2 = \sum_{i=1}^{N_S} S(i)^2 = \frac{( A \cdot V_{FS})^2\cdot ({N_S+3+2/N_S})}{3}\nonumber
\end{equation}

With baseline offset included, each sample is shifted by a constant value $O$. We call this RSS signal $RSS_O$, adding the subscript $O$, and analyse the AC coupled signal over the full period:

\begin{eqnarray}
RSS_O^2 &=& \sum_{i=1}^{N_S+N_B} (S(i)-O)^2\\
    &=& \frac{( A \cdot V_{FS})^2\cdot ({N_S+3+2/N_S})}{3} \nonumber\\
&-& (N_S+N_B) \cdot \left(\frac{(N_S+2)(A V_{FS})}{2\cdot(N_S+N_B)}\right)^2 \nonumber \\
&=& ( A \cdot V_{FS})^2\cdot (N_S+2)  \left( \frac{N_S+1}{3\cdot N_S} - \frac{N_S+2}{4\cdot (N_S+N_B)}\right) \nonumber \\
&\approx&( A \cdot V_{FS})^2\cdot (N_S+2)  \left( \frac{1}{3} - \frac{N_S}{4\cdot (N_S+N_B)}\right) \nonumber \\
&\approx& \frac{( A \cdot V_{FS})^2\cdot (N_S+2)}{3}  \left( 1 - \frac{3\cdot N_S}{4\cdot (N_S+N_B)}\right) \nonumber \\
&\approx& RSS^2 \left( 1 - \frac{3\cdot N_S}{4\cdot (N_S+N_B)}\right)\label{EquRssAcSignal}
\end{eqnarray}
For practical cases of $N_S>>1$ the baseline offset $O$ results in a multiplicative term in the RSS calculation that depends solely on the duty factor of the periodic signal. Note that $N_B$ here indicates the number of baseline samples between pulses. For $N_S=N_B$ the RSS value is reduced to about 65\% and we must expect that this will impact the position uncertainty.

\subsubsection{Position Immunity to AC Coupling}
\label{SecRssAcImmunity}
We now show that the RSS approach delivers the correct estimate also for the case of a baseline offset due to AC coupling. Let us assume that this offset $O$, driven by the beam itself, is also proportional to some effective, integral power $\bar{S}$ of the signal $S$ as was shown in section~\ref{SecTriangleAC} for the triangular pulse:
$O = o \cdot \bar{S}$.

We can then rewrite the RSS position estimator as:
\begin{eqnarray}
\nonumber\left(\frac{<x>}{r}\right)_{RSS}&=&\frac{\sqrt{\Sigma_{i=1}^N{(S_L(i)-O_L)}^2} - \sqrt{\Sigma_{i=1}^N{(S_R(i)-O_R)}^2}}{\sqrt{\Sigma_{i=1}^N{(S_L(i)-O_L)}^2} + \sqrt{\Sigma_{i=1}^N{(S_R(i)-O_R)}^2}} = \\ \nonumber
 &=&\frac{\sqrt{\Sigma_{i=1}^N{(c \cdot S_R(i)- c \cdot o \cdot \overline{S_R})}^2} - \sqrt{\Sigma_{i=1}^N{(S_R(i)- o \cdot \overline{S_R})}^2}}{\sqrt{\Sigma_{i=1}^N{(c \cdot S_R(i)-c \cdot o \cdot \overline{S_R})}^2} + \sqrt{\Sigma_{i=1}^N{(S_R(i)- o \cdot \overline{S_R})}^2}}\\ \nonumber
  &=&\frac{c \cdot \sqrt{\Sigma_{i=1}^N{(S_R(i)- o \cdot \overline{S_R})}^2} - \sqrt{\Sigma_{i=1}^N{(S_R(i)- o \cdot \overline{S_R})}^2}}{c \cdot \sqrt{\Sigma_{i=1}^N{(S_R(i)- o \cdot \overline{S_R})}^2} + \sqrt{\Sigma_{i=1}^N{(S_R(i)- o \cdot \overline{S_R}})^2}}\\
 &=& \frac{c-1}{c+1} 
\end{eqnarray}

Again, we obtain the same value for the position. We stress that this immunity holds only, if the baseline shift is entirely driven by the bunch signal itself. Any other sources of baseline shifts introduce an undetected and unpredictable bias in the position determination.

\subsection{ New Approach: Least-Square Fit of tu\-ples \texorpdfstring{($\Delta,\Sigma$)}{Delta,Sigma}}

\subsubsection{Position Calculation}

A different approach has been proposed in~\cite{3}: a linear regression of the derived quantities difference $\Delta(i)$ versus sum $\Sigma(i)$, where $i$ is the index of the data sample. If the difference data $\Delta(i)$ are analysed as function of the sum data $\Sigma(i)$, one expects the data to be consistent with a straight-line through the origin (direct proportion). Figure~\ref{2D-Fit} illustrates this analysis procedure. The slope parameter is an estimator for the relative position ($\alpha=x/r$) and represents the result of the position measurement:

\begin{equation}\label{Equ2}
\Delta = const \cdot \Sigma = <(x/r)> \Sigma
\end{equation}

{\it This is a direct interpretation of equation~\ref{EquTheory} which states that for a single measurement, i.e. for one given beam position $x$ and sensitivity $S=1/r$, the ratio $\Delta/\Sigma$ is a constant, namely ($x/r$), for each acquired sample.} \\

 In this document the approach of direct proportion is generalised to a straight-line including the intercept term. Both models are extensively treated in the book by R. Barlow~\cite{8} whose notation for estimators and uncertainties we have used as a guideline. Parameter estimates and respective uncertainties are derived from a least-square minimisation which leads to simple analytical formulae. These formulae can be easily analysed in order to judge the robustness of the position estimator. 
 
 The next sections will discuss the characteristics of these estimators and link the RSS approach to the case of direct proportion. Finally, it is shown that the straight-line fit yields the most reliable and robust estimator. Its main advantage in practical applications is the immunity to external offsets to any of the electrode signals (e.g. low frequency amplifier noise or drifts) because they merely displace the origin of the coordinate system without affecting the slope. 

The importance of a baseline-independent "floating signal" analysis for BPM measurements should not be underestimated. To a certain extent it resembles a differential signal transmission which is very robust against common mode interferences. Measurement systems for synchrotron BPMs perform all calculations on-line in an FPGA including the necessary baseline restoration of the raw ADC data. Getting rid of the baseline restoration significantly simplifies the technical implementation and reduces position calculation time~\cite{3}.

\subsubsection{Direct Proportion and Straight-Line}
For the fit procedure the left and right electrode signals are transformed to $(\Delta,\Sigma)$ coordinates. Defined as difference or sum of two variables of equal uncertainty, the uncertainty of the new variables is increased:
\begin{eqnarray}
\sigma_{\Delta} &=& \sqrt{2} \cdot \sigma_{V}\nonumber\\
\sigma_{\Sigma} &=& \sqrt{2} \cdot \sigma_{V}\label{EquUncSample}
\end{eqnarray}

{\it Case 1 - Direct proportion (straight-line through origin):} For a direct proportion $y = m \cdot x$ the following equations hold for the estimator of the slope $<m>$ and its variance $\sigma^2_{<m>}$:

 \begin{eqnarray}
<m> &=& \frac{<x>}{r} = \frac{ \overline{\Delta \cdot\Sigma} }{ \overline{\Sigma^2} }\\
\sigma^2_{<m>} &=& \frac{\sigma_{\Delta}^2}{\sum_{i=1}^{N} \Sigma_i^2} =\frac{\sigma_{\Delta}^2}{N \cdot \overline{\Sigma^2}}\label{EqnSigmaDirect}
\end{eqnarray}
Here $N$ indicates the total number of analysed samples. \\

\begin{figure}[h!]
\begin{center}
\includegraphics[width = 0.9\textwidth]{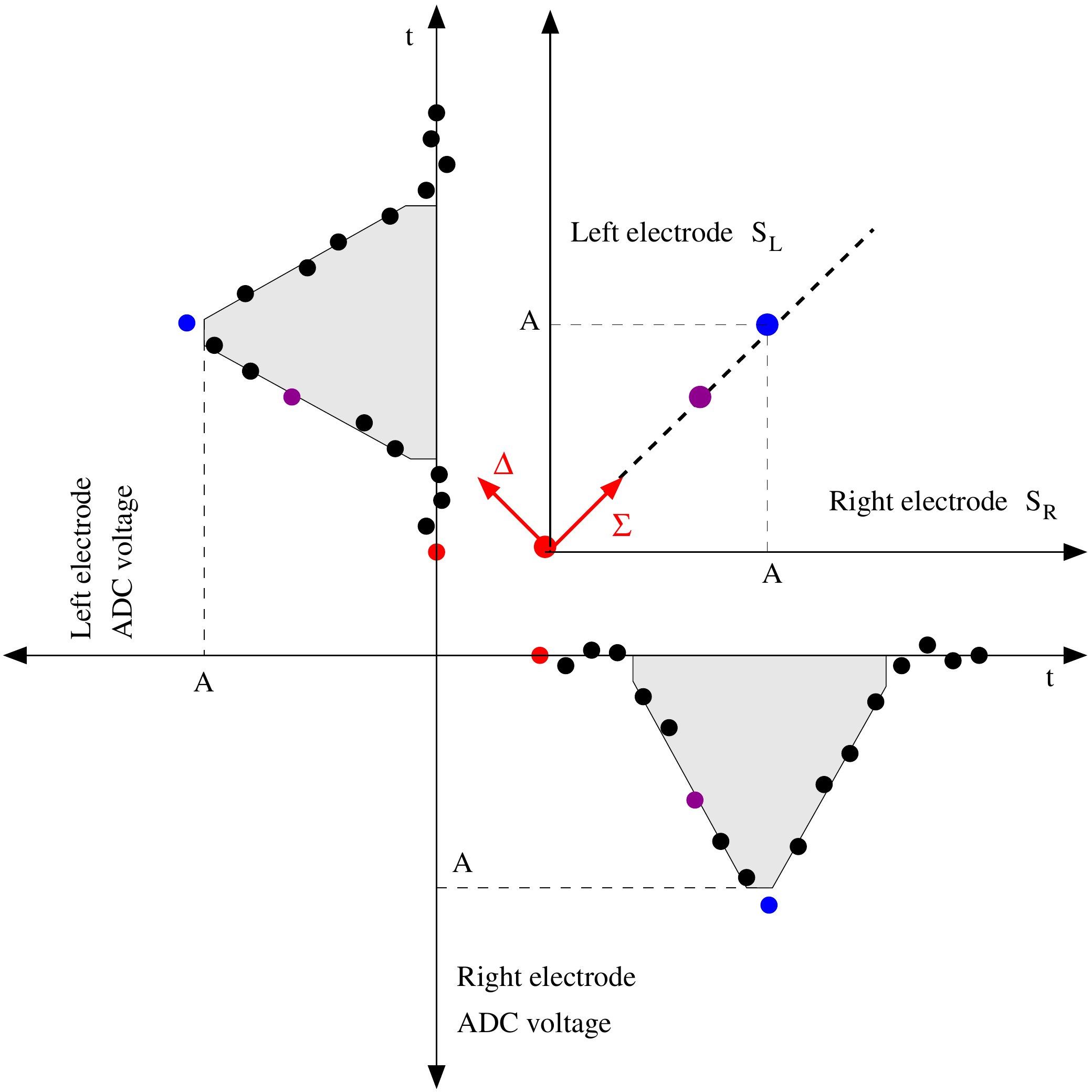}
\caption{Illustration of the fit approach of left against right electrode signal for three data samples. The tuple \{$(1/\sqrt{2})(1,-1),(1/\sqrt{2})(1,1)$\} represents a new basis of the two-dimensional vector space. It is generated from the canonical basis \{(1,0),(0,1)\} by a 45$^\circ$ rotation and defines the ($\Delta$,$\Sigma$) coordinates.}
\label{2D-Fit}
\end{center}
\end{figure}

{\it Case 2 - Straight-line fit:} The least-square minimisation for a straight-line $y = m\cdot x + c$ leads to the following estimators for slope $<m>$, intercept $<c>$, their variances and the covariance:
 \begin{eqnarray}
<m> &=&  \frac{<x>}{r} = \frac{cov(\Delta, \Sigma)}{\sigma^2_\Sigma} \\
\sigma^2_{<m>} &=& \frac{\sigma_{\Delta}^2}{N \cdot \sigma^2_\Sigma}\label{EquVarLine}\\
<c> &=& \overline{\Delta} - <m> \cdot \overline{\Sigma}\\
\sigma^2_{<c>} &=& \overline{\Sigma^2} \cdot \sigma^2_{<m>}\\
cov(m,c) &=& -\overline{\Sigma}\cdot \sigma^2_{<m>}
\end{eqnarray}
Please note that the derivation of these equations assumes error bars in the measured quantity only (here: $\Delta$), disregarding the uncertainty at the measurement value (here: $\Sigma$). This is not true in the present case since both variables carry the same uncertainty. The problem is often dealt with in a further iteration step after the initial fit by introducing the {\it effective variance} $\sigma_{eff}^2= \sigma_{\Delta}^2+ <m>\sigma_{\Sigma}^2$ and repeating the fit procedure with this uncertainty. The uncertainty of the horizontal $\Sigma$ coordinate is simply transferred to the vertical $\Delta$ axis. 

In our application however, $<m>\,\sim 0$ for a beam around the reference orbit and we are left with the original "vertical" component of the uncertainty. Had we chosen to fit $S_L$ versus $S_R$, the expectation value $<m>\,\sim 1$ and this would have to be included in the effective variance. Note that in defining $\Delta$ and $\Sigma$ we have introduced a basis transformation equivalent to a 45$^\circ$ rotation as shown in Figure~\ref{2D-Fit}.\\

{ \it The fit approach differs fundamentally from the classical integral approach, since it estimates the slope parameter ($x/r$) from the two-dimensional tuples of derived, coupled quantities ($\Delta$,$\Sigma$), rather than calculating this parameter from the two independent integral estimates of the one-dimensional electrode signals $S_L$ and $S_R$. The measurement observable is the correlation between the two signals. The integral approach requires knowledge of the baseline to calculate correct integrals, while the fit approach does not!}

\subsubsection{Discussion of Direct Proportion}
We now state two simple and perhaps trivial, yet very instructive equivalences.\\
 
{\it 1) Equivalence to Weighted Average}\\

Our first statement is quite intuitive: The estimator $<m>$ is equivalent to a weighted average of single position measurements $m_i$  derived from each tuple $(\Delta_i,\Sigma_i)$.
\begin{eqnarray}
m_i &=& \Delta_i/\Sigma_i \\
<m> &=& \frac{(\sum_{i=1}^{N} \Sigma_i \cdot \Delta_i)/N}{ (\sum_{i=1}^{N} \Sigma^2_i)/N} = \frac{\sum_{i} (\Sigma_i \cdot \Delta_i)}{\sum_{i} \Sigma^2_i} = \frac{\sum_{i} (\Sigma^2_i \cdot (\Delta_i/\Sigma_i))}{\sum_{i} \Sigma^2_i} \\ \nonumber
   &=& \frac{\sum_{i} (\Sigma^2_i \cdot m_i)}{\sum_{i} \Sigma^2_i} = \frac{\sum_{i} (w_i \cdot m_i)}{\sum_{i} w_i}
\end{eqnarray}

We have arrived at the well known equation for the weighted average and have repeated the "direct" interpretation of equation~\ref{EquTheory}: Each tuple $(\Delta_i,\Sigma_i)$ conveys information on the beam position. The weight $w_i(m_i)$ of a single measurement $m_i$ (or tuple) is given by $w_i(m_i)=\Sigma_i^2$, in other words, its uncertainty $\sigma(m_i) = 1/\Sigma_i$. We could have also chosen to normalise the weights $w_i$ by $RSS^2$ as $w_i(m_i)=\Sigma_i^2/RSS^2$ (which is equivalent to equ.~\ref{EqnWeightRSS}). 
The coordinate origin is assumed to be known which seems not justified for an AC coupled system: The amplitude $\Sigma_i$ changes as the signal baseline drifts, e.g. during a synchrotron cycle.\\

{\it 2) Equivalence to RSS estimator}\\

 The second statement might not be so obvious: The RSS approach is equivalent to the direct proportion fit, if we make use of equation~\ref{EquSLSR} again, $S_L = c \cdot S_R$. Then, we can rewrite the estimator $<m>$:

\begin{eqnarray}
<m> &=& \frac{<x>}{r} = \frac{ \overline{\Delta \cdot\Sigma} }{ \overline{\Sigma^2} }\\ \nonumber
    &=& \frac{ \sum_i [S_L(i)-S_R(i) ][ S_L(i)+S_R(i) ] } {\sum_i [ S_L(i)+S_R(i) ]^2}\\ \nonumber
    &=& \frac{ \sum_i [c \cdot S_R(i)-S_R(i) ][ c \cdot S_R(i)+S_R(i) ] }{ \sum_i [ c \cdot S_R(i)+S_R(i)]^2 }\\ \nonumber
    &=& \frac{\sum_i (c-1)(c+1) \cdot S_R^2(i)}{\sum_i (c+1)^2 \cdot S_R^2(i)}\\ \nonumber
    &=& \frac{(c-1)(c+1) \cdot \sum_i S_R^2(i)}{(c+1)^2 \cdot \sum_i S_R^2(i)}=\frac{c-1}{c+1}
\end{eqnarray}

The estimator $<m>$ yields the same result that was derived in section~\ref{SecRss=Int}. Hence, RSS approach and direct proportion extract the same beam position from the data.

\subsubsection{Discussion of Straight-Line}

\begin{figure}[h]
\begin{center}
\includegraphics[width = 70mm]{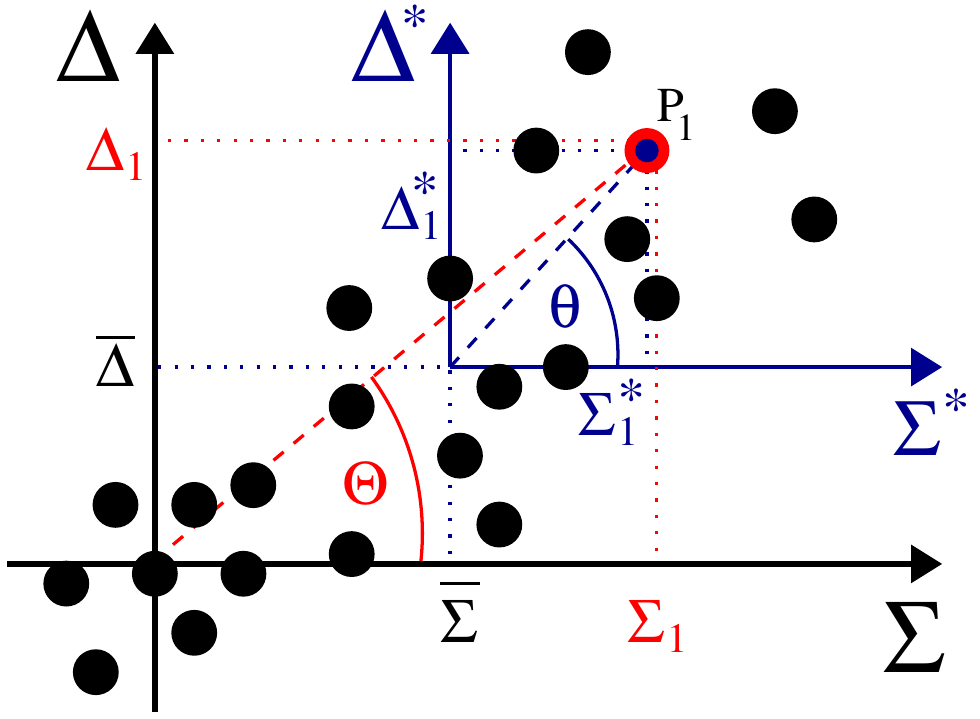}
\caption{Illustration of weighted average calculation. Dots represent the distribution of $(\Sigma, \Delta)$ tuples. Each tuple represents a single position measurement. This is illustrated for sample P$_\mathrm{1}$ marked in red and blue colour. The direct proportion takes reference to the black coordinate system centered at (0/0) and assigns a weight $w_1=\Sigma_1^2$, while the straight-line fit references to the blue system centered at ($\overline{\Sigma}$, $\overline{\Delta}$) and assigns weight $w_1^\ast=(\Sigma_1^\ast)^2$.}
\label{WeightedMean}
\end{center}
\end{figure}

{\it 1) Equivalence to Weighted Average}\\
We repeat the calculation in the same manner as for the direct proportion and arrive at a similar result: The straight-line fit is equivalent to a weighted mean of position measurements
$m_i = (\Delta_i- \overline{\Delta})/(\Sigma_i-\overline{\Sigma})$, weighted by $w_i= (\Sigma_i-\overline{\Sigma})^2$. Again, one could have chosen to normalise the weights. The assigned weights are illustrated in Figure.~\ref{WeightedMean}.\\

{\it 2) Equivalence to RSS estimator}\\
The estimator $<m>$ for the straight-line yields the same result that was derived in section~\ref{SecRss=Int}. Hence, Integral approach, RSS approach, direct proportion extract the same beam position from the data.\\

{\it 3) Robustness of Estimator}\\
So far we have repeatedly made use of equation~\ref{EquSLSR}, $S_L = c \cdot S_R$, in the treatment of the data analysis. But one might ask, what happens if this simple relation is violated? Clearly, all our previous results would be affected, most even invalid. \\

In practical applications one would like to work with an estimator that is immune, at least, to an additional offset such that $S_L = c_1 \cdot S_R + c_0$. This offset $c_0$ could be an ADC offset or a low frequency amplifier disturbance or noise pickup which mimics a constant offset in a short measurement window. For different BPMs along a ring there might be different noise contributions to $S_L$ and $S_R$ along the signal transmission in the electronics chains, too. All of these effects are very elusive in day-to-day operation and introduce position biases that are difficult to quantify when beam positions, global orbit or tune values are to be analysed.\\

{\it The answer to the posed question is simple: Fit a straight-line, rather than a direct proportion, to the data. It is immediately obvious from the structure of the slope estimator which is defined by the ratio of covariance $cov(\Sigma, \Delta)$ and variance $\sigma^2_{\Sigma}$. 
Both quantities intrinsically refer to the mean values of $\overline{\Delta}$ and $\overline{\Sigma}$. 
Therefore, the estimator is independent of the actual value of the tuple mean $( \overline{\Sigma}, \overline{\Delta} )$ and is able to adjust to the "floating" origin of the coordinate system in the AC coupled electronics. 
In other words: Moving the coordinate system does not change the slope. This is not the case, when the line is forced through the origin of the coordinate system as in the case of direct proportion.}

\section{Calculation of Position Uncertainty}
\label{SecCalcUncertainty}
We have shown that all approaches result in the same position estimate under ideal conditions. In the next three sections, one for each approach, we proceed with the calculation of the position uncertainty for both pulse shapes assuming a centered beam. In the final section we summarise the most important results. The discussion emphasis is placed on the triangular pulse shape since it represents the more realistic case.

\subsection{Classical Approach 1 - Signal Integration}
\label{SecUncertaintyInt}
We can calculate the position uncertainty for independent data samples from equation~\ref{Equ1} in a straight-forward manner:
\begin{equation}\nonumber
\frac{<x>}{ r}= \frac{I_L - I_R}{I_L + I_R}
\end{equation}

The BPM radius $r$ is constant and error propagation for the integral variables $I_L$ and $I_R$ leads to:
\begin{equation}
\frac{\sigma_{ <x>}}{r} =  \frac{2}{(I_L + I_R)^2}\cdot \sqrt{(I_R \cdot\sigma_{I_L})^2 + (I_L \cdot\sigma_{I_R})^2}
\end{equation}

Assuming a centred beam and hence $I_L$=$I_R$, the uncertainty is given by:
\begin{eqnarray}
\nonumber \frac{\sigma_{<x>}}{r} &=&  \frac{2}{(2 I_L)^2}\cdot \sqrt{(I_L \cdot\sigma_{I_L})^2 + (I_L \cdot\sigma_{I_L})^2}\\
\frac{\sigma_{<x>}}{r} &=&  \frac{2}{(2 I_L)^2}\cdot \sqrt{2}\cdot I_L \sigma_{I_L} =  \frac{\sigma_{I_L}}{\sqrt{2}\cdot I_L}\label{EqnClassic1}
\end{eqnarray}

\subsubsection{Square Pulse}
If the beam is centred, the signal level S is identical for both electrodes. Using the square pulse definitions of section~\ref{SecSquare} we obtain for the integral variables omitting the index L/R for the electrode:
\begin{eqnarray}
I &=& N_S\cdot A\cdot V_{FS} \\
\sigma_{I} &=& \sigma_V \cdot  \sqrt{N_S}   \textrm{\quad for signal samples only ($N_B$=0)}\\
\sigma_{I} &=& \sigma_V \cdot  \sqrt{N_S+N_B} \textrm{\quad for all samples ($N_B>$0)}
\end{eqnarray}
The integral assigns identical weights $w=(dI/dS(i))^2=1$ to the data and does not prefer signal over baseline samples. 

Finally, for a centered beam we set A=$A_L$=$A_R$ and rewrite equation~\ref{EqnClassic1}:
\begin{eqnarray}
\textrm{$N_B$=0\,:\quad} \frac{\sigma_{<x>}}{ r} &=&  \frac{1}{\sqrt{2}} \cdot \big(\frac{\sigma_V}{A\cdot V_{FS}}\big)\cdot \frac{1}{\sqrt{N_S}}\label{EqnSquInt} \\
\textrm{$N_B>$0\,:\quad}\frac{\sigma_{<x>}}{ r} &=&   \frac{1}{\sqrt{2}} \cdot  \big(\frac{\sigma_V}{A\cdot V_{FS}}\big) \cdot \frac{\sqrt{N_S+N_B}}{ N_S}\label{EqnSquInt2}
\end{eqnarray}

We have arrived at an intuitive result: Our knowledge of the measured position improves for small voltage jitter and when the signal level $A$ is close to the full scale value $V_{FS}$. Further, the measurement improves, if baseline samples $N_B$ can be excluded without cutting into the signal samples $N_S$. In other words, after baseline restoration the baseline is considered as a noise source.

For off-centred beams, the two signal amplitudes $A_L$ and $A_R$ are not equal, and equation~\ref{EqnSquInt2} reads:
\begin{eqnarray}
\textrm{$N_B>$0\,:\quad}\frac{\sigma_{<x>}}{ r} &=&   2 \cdot  \frac{\sigma_V}{ V_{FS}} \cdot \frac{\sqrt{A_L^2 + A_R^2}}{(A_L + A_R)^2} \cdot \frac{\sqrt{N_S+N_B}}{ N_S}\label{EqnSquInt3}
\end{eqnarray}

\subsubsection{Triangular Pulse}
We continue the discussion for the case of a triangular pulse which seems a better representative of a beam pulse. Using the definitions of section~\ref{SecTriangle} we obtain for the integral variables omitting the electrode index $L/R$:
\begin{eqnarray}
I &=& \sum_{i=1}^{N_S} S(i) =  2\sum_{i=1}^{N_S/2} S(i) =\frac{N_S+2}{2} \cdot A\cdot V_{FS} \label{EquTriInt}\\
\sigma_{I} &=& \sigma_V \cdot  \sqrt{N_S}   \textrm{\quad for signal samples only ($N_B$=0)}\\
\sigma_{I} &=& \sigma_V \cdot  \sqrt{N_S+N_B} \textrm{\quad for all samples ($N_B>$0)}\label{EquTriIntError}
\end{eqnarray}

Thereby, we have used the relation $\sum_i^n i = \frac{1}{2}n(n+1)$ and arrive at the well known formula for a triangle area of total width ($N_S+2$). 
For a centered beam we set A=$A_L$=$A_R$ and rewrite equation~\ref{EqnClassic1}:
\begin{eqnarray}
\boxed{
\begin{aligned}
\textrm{$N_B$=0\,:\quad} \frac{\sigma_{<x>}}{ r} &=  \sqrt{2} \cdot \big(\frac{\sigma_V}{A\cdot V_{FS}}\big)\cdot \frac{\sqrt{N_S}}{N_S+2}\label{EqnTriInt} \\
\textrm{$N_B>$0\,:\quad}\frac{\sigma_{<x>}}{ r} &=   \sqrt{2} \cdot  \big(\frac{\sigma_V}{A\cdot V_{FS}}\big) \cdot \frac{\sqrt{N_S+N_B}}{ N_S+2}\label{EqnTriInt2}
\end{aligned}
}
\end{eqnarray}

For large $N_S$ we arrive at the same formula as for the square pulse, but for the smaller effective amplitude $A_{eff}=\frac{1}{2} A$ due to the smaller area of the triangular shape.

For off-centred beams, the two signal amplitudes $A_L$ and $A_R$ are not equal, and equation~\ref{EqnTriInt2} reads:
\begin{eqnarray}
\boxed{
\begin{aligned}
\textrm{$N_B>$0\,:\quad}\frac{\sigma_{<x>}}{ r} &=&   4 \cdot  \frac{\sigma_V}{ V_{FS}} \cdot \frac{\sqrt{A_L^2 + A_R^2}}{(A_L + A_R)^2} \cdot \frac{\sqrt{N_S+N_B}}{ N_S+2}\label{EqnTriInt3}
\end{aligned}
}
\end{eqnarray}
\subsubsection{Uncertainty due to Baseline Restoration}
\label{SecUncertaintyBLR}
If the baseline offset $O$ is estimated as mean value  $<O>$ from $N_O$ samples outside the position calculation region (which includes $N_S$ signal and $N_B$ baseline samples), the uncertainty $\sigma_{O}$ is given as
\begin{equation}
\sigma_{<O>} = \frac{\sigma_V}{\sqrt{N_O}}
\end{equation}
This value if subtracted from all samples that are part of the position calculation and this step introduces a common correlation (see~\cite{8}, chapter 4). The uncertainty of the integral $I_{corr}$ over the baseline-corrected data is then given by

\begin{eqnarray}
\sigma_{I_{corr}} &=& \sigma_V \cdot  \sqrt{N_S + \frac{N_S^2}{N_O}}   \textrm{\quad for signal samples only ($N_B$=0)}\\
\sigma_{I_{corr}} &=& \sigma_V \cdot  \sqrt{(N_S+N_B) + \frac{(N_S+N_B)^2}{N_O}} \textrm{\quad for all samples ($N_B>$0)}\label{EquUncTriIntError}
\end{eqnarray}
The correlation introduces an additional term proportional to the total number of samples $N^2$. Hence, it is important to exclude baseline samples $N_B$ from the integral calculation and to include a maximum number of baseline samples $N_O$ in the calculation of the offset value $O$; not always easy tasks if one looks at real bunch signals.

\subsection{Classical Approach 2 - RSS Calculation}
\label{SecUncertaintyRSS}
We can calculate the position uncertainty for independent data samples from equation~\ref{Equ1} in a straight-forward manner.
\begin{equation}
\frac{<x>}{ r}= \frac{RSS_L - RSS_R}{RSS_L + RSS_R}
\end{equation}

Here $RSS_L/R$ are given by the root-sum-square RSS value defined as:
\begin{align}\label{eq:pow_dev}
RSS_{L/R} &= \sqrt{\sum\limits_{i=1}^{N} {S_{L/R}(i)}^2} \textrm{\quad and}\\
\sigma_{RSS} &= {\sigma_V}
\end{align}

The uncertainty of the RSS value is independent of the sample number and is simply given by the uncertainty of the single ADC sample $\sigma_V$. The reasons is that the RSS method suppresses noise contributions from baseline samples since it assigns proportional weights $w(i)$ to samples according to their signal strength:
\begin{equation}
w(i)=\left(\frac{dRSS}{dS(i)}\right)^2= \left(\frac{S(i)}{RSS}\right)^2 \label{EqnWeightRSS}
\end{equation}
For the position estimator error propagation for the RSS variables $RSS_L$ and $RSS_R$ leads to:
\begin{equation}
\frac{\sigma_{ <x>}}{r} =  \frac{2}{(RSS_L + RSS_R)^2}\cdot \sqrt{(RSS_R \cdot\sigma_{RSS_L})^2 + (RSS_L \cdot\sigma_{RSS_R})^2}
\end{equation}

Assuming a centred beam and hence $RSS$=$RSS_L$=$RSS_R$, the uncertainty is given by:
\begin{eqnarray}
\nonumber\frac{\sigma_{<x>}}{r} &=&  \frac{2}{(2 \cdot RSS_L)^2}\cdot \sqrt{(RSS_L \cdot\sigma_{RSS_L})^2 + (RSS_L \cdot\sigma_{RSS_L})^2}\\
\nonumber\frac{\sigma_{<x>}}{r} &=&  \frac{2}{(2 \cdot RSS)^2}\cdot \sqrt{2}\cdot RSS \cdot \sigma_{RSS} =  \frac{\sigma_{RSS}}{\sqrt{2}\cdot RSS}\\
\frac{\sigma_{<x>}}{r} &=& \frac{\sigma_V}{\sqrt{2} \cdot RSS}\label{EqnClassic2}
\end{eqnarray}

\subsubsection{Square Pulse without Baseline Offset}

We start with the simple case of a square pulse without baseline offset (see section~\ref{SecSquare}) and use equation~\ref{EqnClassic2}.

\begin{eqnarray}
RSS^2 &=& \sum_{i=1}^{N_S+N_B} S(i)^2 = N_S \cdot (A \cdot V_{FS})^2 \nonumber\\
\frac{\sigma_{<x>}}{r} &=& \frac{1}{\sqrt{2}}\cdot (\frac{\sigma_V}{A \cdot V_{FS}}) \cdot \frac{{1}}{\sqrt{N_S}}
\label{EqnUncRssSquNoOff}
\end{eqnarray}
The baseline samples $N_B$ do not convey any information and therefore the uncertainty only depends on the number of signal samples $N_S$. The result is identical to equation~\ref{EqnSquInt} for the integration approach without baseline samples $N_B$.

\subsubsection{Square Pulse with Baseline Offset}

We continue with the case of a square pulse with baseline offset $O$ (see section~\ref{SecSquareAC}) and use equations~\ref{EquOffSqu} and~\ref{EqnClassic2}. Note that $N_B$ is the number of samples between two successive pulses. It is not the number of analysed samples outside the signal region. This case treats the turn-by-turn analysis for harmonic $h=1$.

\begin{eqnarray}
RSS_O^2 &=& \sum_{i=1}^{N_S+N_B} (S(i)-O)^2 = N_S \cdot (A \cdot V_{FS})^2 \cdot \frac{N_B}{N_S+N_B} \nonumber \\
\frac{\sigma_{<x>}}{r} &=&\frac{1}{\sqrt{2}}\cdot (\frac{\sigma_V}{A \cdot V_{FS}})  \sqrt{\frac{N_S+N_B}{N_S \cdot N_B} }
\end{eqnarray}

\subsubsection{Square Pulse: general case}
Finally, we may not include all samples $N_B$ between pulses, but a smaller number $N_b<N_B$, and consider an off-centre beam. Then the uncertainty is given by:
\begin{equation}
\boxed{
\frac{\sigma_{<x>}}{r} = 2 \cdot \frac{\sigma_V}{V_{FS}} \cdot \frac{\sqrt{A_L^2 + A_R^2}}{(A_L+A_R)^2} \cdot \frac{(N_S+N_B)}{\sqrt{N_S(N_b \cdot N_S+N_B^2)} }
}
\label{EquUncRssSquGen}
\end{equation}

\subsubsection{Triangular Pulse without Baseline Offset}
\label{SecUncRssTri}
We continue the discussion for the case of a triangular pulse which seems a better representative of a beam pulse. Using the definitions of section~\ref{SecTriangle} we obtain for the RSS variables where we have omitted the electrode index $L/R$:
\begin{eqnarray}
RSS^2 &=& 2\cdot \sum_{i=1}^{N_S/2} S(i)^2 = \frac{( A \cdot V_{FS})^2\cdot ({N_S+3+2/N_S})}{3} \label{EquRSS2}
\end{eqnarray}\\
\begin{equation}
\boxed{
\frac{\sigma_{<x>}}{r} =\sqrt{3/2}\cdot (\frac{\sigma_V}{A \cdot V_{FS}}) \cdot \frac{{1}}{\sqrt{({N_S+3+2/N_S})}}
}
\label{Eq:urror}
\end{equation}

Note that the position uncertainty is independent of the number of baseline samples $N_B$ as these samples are assigned a weight of zero. The uncertainty reduces only with the number of signal samples $N_S$.

\subsubsection{Triangular Pulse with Baseline Offset}
\label{SecUncRssTriOffset}
Now we consider the case of a triangular pulse with baseline offset $O$ and treat the case of turn-by-turn analysis for harmonic $h=1$. Using the results of section~\ref{SecRssAcSignal}, namely equation~\ref{EquRssAcSignal} for $N_S>>1$, we obtain:
\begin{eqnarray}
RSS_O^2 &\approx& RSS^2  \left( 1 - \frac{3\cdot N_S}{4\cdot(N_S+N_B)} \right) \label{EquRssOffset}
\end{eqnarray}

\begin{equation}
\boxed{
\frac{\sigma_{<x>}}{r} \approx \sqrt{3/2}\cdot (\frac{\sigma_V}{A \cdot V_{FS}}) \cdot \frac{{1}}{\sqrt{({N_S+3+2/N_S})}} \cdot \frac{1}{\sqrt{1-\frac{3}{4}\cdot \frac{N_S}{N_S+N_B}}}
}
\label{EquUncRssTriAC}
\end{equation}
\begin{eqnarray}
\frac{\sigma_{<x>}}{r} \approx \sqrt{\frac{3}{2}}\cdot (\frac{\sigma_V}{A \cdot V_{FS}})  \cdot \sqrt{\frac{N_S\cdot (N_S+N_B)}{(N_S+2)(\frac{1}{4} N_S^2+N_S\cdot N_B+N_B +\frac{1}{4}N_s)}}\nonumber
\end{eqnarray}

The uncertainty depends on the duty factor of the signal and therefore the number of samples $N_B$ which fill the region between two successive pulses. When the straight-line fit is discussed, we will find the second form of the equation again (see equation~\ref{EquTriSigma}), if the leading terms are kept. Therefore, direct proportion and straight-line fit yield the same estimator and uncertainty for the position.
 
\subsubsection{Triangular Pulse: general case}
\label{SecUncRssTriOffsetGeneral}
Finally, for off-center beams and $N_b$ baseline samples around the signal we obtain:
\begin{eqnarray}
\frac{\sigma_{<x>}}{r} &\approx&  \frac{  2 \sqrt{3}  \cdot\sigma_V}{ V_{FS}} \cdot \frac{\sqrt{A_L^2 + A_R^2}}{(A_L + A_R)^2}\nonumber
\\ &&\cdot \frac{{1}}{\sqrt{({N_S+3+\frac{2}{N_s}})(1-\frac{3}{4} \frac{N_S}{N_S+N_B}\cdot (1 - \frac{N_b-N_B}{N_S+N_B}))}} \nonumber
\end{eqnarray}


\subsection{Least-Square Fit Approach}
\label{SecUncertaintyFit}
We briefly repeat the basic formulae for the fit estimators as given in reference~\cite{8}. These equations are now solved for the two signal shapes, and we focus on the calculation of the position uncertainty.\\

{\it Case 1 - Direct proportion (straight-line through origin):} For a direct proportion the following equations hold for the estimator of the slope $m$ and its uncertainty:
 \begin{eqnarray}
<m> &=& \frac{<x>}{r} = \frac{ \overline{\Delta \cdot\Sigma} }{ \overline{\Sigma^2} }\\
\sigma^2_{<m>} &=& \frac{\sigma_{\Delta}^2}{\sum_{i=1}^{N} \Sigma_i^2} =\frac{\sigma_{\Delta}^2}{N \overline{\Sigma^2}}
\end{eqnarray}
Here we have defined the total number of samples $N =N_S+N_B$. \\

{\it Case 2 - Straight-line fit:} The least-square minimisation of a straight-line leads to the following estimators for slope $m$ and its uncertainty:
 \begin{eqnarray}
<m> &=&  \frac{<x>}{r} = \frac{cov(\Delta, \Sigma)}{\sigma^2_\Sigma} \\
\sigma^2_{<m>} &=& \frac{\sigma_{\Delta}^2}{N \cdot \sigma^2_\Sigma}\label{EqnSigmaStraight}
\end{eqnarray}

\subsubsection{Square Pulse}

{\it Case 1 - Direct Proportion \& no baseline offset:}
 \begin{eqnarray}
<m> &=& \frac{cov(\Delta, \Sigma)}{\sigma^2_\Sigma} = \overline{\Delta} / \overline{\Sigma} \label{EqnSquCase1} \\
\sigma^2_{<m>} &=& \frac{\sigma_{\Delta}^2}{ \sum_{i=1}^N \Sigma_i^2}
\end{eqnarray}

We need to calculate the denominator of equation~\ref{EqnSigmaDirect}:\\
\begin{equation}
N\overline{\Sigma^2} = \sum_{i=1}^{N} \Sigma_i^2 = N_S \cdot(2\cdot A \cdot V_{FS})^2
\end{equation} 

Finally, for a centered beam we set A=$A_L$=$A_R$ and rewrite equation~\ref{EqnSigmaDirect} inserting equation~\ref{EquUncSample} for the sample uncertainty:
\begin{equation}
\sigma_{<m>} = \frac{\sigma_{<x>}}{ r} =  \frac{1}{\sqrt{2}} \cdot \big(\frac{\sigma_V}{A\cdot V_{FS}}\big)\cdot \frac{1}{\sqrt{N_S}}\label{EqnSigSquCase1}
\end{equation}
Note that the uncertainty is almost independent of the number of background samples $N_B$ as they do not contribute significantly to the denominator of equation~\ref{EqnSigmaDirect}:
\begin{equation}
 \sum_{i=1}^{N} \Sigma_i^2 = N_S \cdot(2\cdot A \cdot V_{FS})^2 + N_B\cdot \sigma_{\Delta}^2
\end{equation}
After all, we have assumed very good knowledge of the origin. Equation~\ref{EqnSigSquCase1} is identical to equation~\ref{EqnSquInt} for the classical approach and equation~\ref{EqnUncRssSquNoOff} for the direct proportion.\\

For completeness we derive equation~\ref{EqnSigSquCase1} using the propagation of errors of the mean values in equation~\ref{EqnSquCase1} for the case of a centered beam:
\begin{equation}
\sigma^2_{<m>} = \frac{\sigma^2_{\overline{\Delta}}}{(\overline{\Sigma})^2} +\left(\frac{\overline{\Delta}\cdot \sigma_{\overline{\Sigma}}}{(\overline{\Sigma})^2}\right)^2 
\end{equation}
For a centered beam the second part vanished ($\overline{\Delta}=0$) and we obtain for a measurement without baseline samples:
\begin{equation}
\sigma_{<m>} = \frac{\sigma_{<x>}}{ r} =\frac{\sigma_{\overline{\Delta}}}{(\overline{\Sigma})} = \frac{\sqrt{2} \sigma_V / \sqrt{N_S}}{2 \cdot A \cdot V_{FS}} =  \frac{1}{\sqrt{2}} \cdot  \big(\frac{\sigma_V}{A\cdot V_{FS}}\big)\cdot \frac{1}{\sqrt{N_S}}\label{EquSig1}
\end{equation}
We have used $\sigma_{\overline{\Delta}} = \sigma_{\Delta}/\sqrt{N_S}  = \sqrt{2}\sigma_V/\sqrt{N_S}$ and obtain again the previous result. Baseline samples can be included by calculating the effective amplitude $\overline{\Sigma_{eff}}$:
\begin{equation}
 \overline{\Sigma_{eff}} =  \overline{\Sigma}\cdot\frac{N_S}{N_S+N_B}
\end{equation}
Inserting $\overline{\Sigma_{eff}}$ into equation~\ref{EquSig1} and substituting $N_S$ with $N_S+N_B$ leads to equation~\ref{EqnSquInt2}.\\

In order to estimate the influence of the 2nd term, we insert equation~\ref{Equ2} in the equation and use $\sigma_{\Delta}=\sigma_{\Sigma}$ as well as $\sigma_{\overline{\Delta}}=\sigma_{\overline{\Sigma}}$:
\begin{equation}
\sigma^2_{<m>} = \frac{\sigma^2_{\overline{\Delta}}}{(\overline{\Sigma})^2} +(\frac{\overline{\Delta}\cdot \sigma_{\overline{\Sigma}}}{(\overline{\Sigma})^2})^2 = \frac{\sigma^2_{\overline{\Delta}}}{(\overline{\Sigma})^2}\cdot \left[\,1+ (<x>/r)^2 \right]
\end{equation}
Since the additive term is of higher order, the uncertainty growth is small to moderate: for small offsets $(x/r)<0.1$ the uncertainty is increased by less than 1\%, for a significant offset of $(x/r)=0.3$, the increase is below 10\%. We may disregard this term for practical applications altogether.\\

{\it Case 2 - Straight-line fit:} In a practical application we have no exact knowledge of the two signal baselines and hence the origin of the two-dimensional coordinate space $(\Delta, \Sigma)$. Therefore, we have to add another free parameter and fit a straight-line through the data. The position uncertainty is now given by equation~\ref{EqnSigmaStraight}:
 \begin{equation}
\sigma_{<m>} = \frac{\sigma_{\Delta}}{\sqrt{N} \cdot \sigma_\Sigma}
\end{equation}

Therefore we have to calculate the standard deviation of $\Sigma$ and start with the well known equation:
 \begin{eqnarray}
\sigma^2_{\Sigma} &=& \overline{\Sigma^2}-  \overline{\Sigma} ^2 \\
\overline{\Sigma} &=& \frac{2\cdot A \cdot V_{FS} \cdot N_S}{N_S+N_B}\\
\overline{\Sigma^2} &=& \frac{1}{N_S+N_B} \cdot \big(N_S \cdot (2\cdot A \cdot V_{FS}) ^2+ N_B\cdot \sigma_{\Delta}^2\big)\label{EquSquSig2}
\end{eqnarray}
Finally we obtain, if the background contribution in the second term is disregarded in equation~\ref{EquSquSig2}:
\begin{eqnarray}
\sigma^2_{\Sigma} &=& \frac{N_S\cdot N_B}{(N_S+N_B)^2} \cdot (2\cdot A \cdot V_{FS}) ^2\\
\frac{\sigma_{<x>}}{r}   &=& \frac{1}{\sqrt{2}}\cdot \big(\frac{\sigma_V}{A \cdot V_{FS}}\big) \cdot \frac{\sqrt{N_S+N_B}}{\sqrt{N_S \cdot N_B}} 
\end{eqnarray}
The result is symmetric with respect to $N_S$ and $N_B$ and similar to equation~\ref{EqnSquInt2}. But now the term $\sqrt {N_S\cdot N_B}$ replaces the signal sample number $N_S$. This is explained by the fact, that we have given up the fixed reference of an origin or zero baseline level! The square pulse produces two data points in the $(\Delta,\Sigma)$ coordinate system and this case is equivalent to a fit through two points. Better knowledge on any of the positions, i.e. a larger number of samples at either end, must improve the outcome of the result. For  $N_B=N_S$ we obtain the same results as in the classical case, namely equations~\ref{EqnSquInt2}.

\subsubsection{Triangular Pulse}

{\it Case 1 - Direct Proportion \& no baseline offset:}
 \begin{eqnarray}
\sigma^2_{<m>} &=& \frac{\sigma_{\Delta}^2}{ \sum_{i=1}^N \Sigma_i^2}\\
                          &=& \frac{\sigma_{\Delta}^2}{2\cdot \sum_{i=1}^{N_2/2} (\frac{4\cdot A}{N_S} \cdot V_{FS} \cdot i)^2}=  \frac{\sigma_{\Delta}^2}{2 (\frac{4\cdot A}{N_S} V_{FS})^2  \sum_{i=1}^{N_2/2} i}\nonumber\\
                        &=& \frac{6\cdot\sigma_{\Delta}^2}{2 (\frac{4\cdot A}{N_S} V_{FS})^2 \cdot N_S/2 \cdot (N_S+2)/2 \cdot (N_S+1)}\nonumber\\
                       &=& \frac{3/4 \cdot \sigma_{\Delta}^2}{(A \cdot V_{FS})^2}\cdot \frac{1}{N_S+3+2/N_S}\nonumber
\end{eqnarray}
\begin{equation}
\boxed{
\frac{\sigma_{<x>}}{r} =\sqrt{\frac{3}{2}}\cdot \big(\frac{\sigma_{V}}{A \cdot V_{FS}}\big)\cdot \frac{1}{\sqrt{N_S+3+2/N_S}}
}\label{EquTriFitSigma}
\end{equation}
We have derived equation~\ref{Eq:urror} again which is expected from the equivalence between RSS approach and direct proportion fit.\\

{\it Case 2 - Straight-line fit:}
In a last step we calculate the uncertainty for a triangular pulse for a straight-line fit as given in equation~\ref{EquVarLine}. First we need to calculate the denominator:

\begin{eqnarray}\nonumber
N\cdot \sigma^2_{\Sigma} &=&  N\cdot (\overline{\Sigma^2} - (\overline\Sigma)^2) \nonumber\\
                                        &=& \sum_{i=1}^N \Sigma^2_i - N\cdot (\overline\Sigma)^2\nonumber\\
	&=& \frac{4}{3}\cdot \frac{(A \cdot V_{FS})^2}{N_S}\cdot (N_S+1)(N_S+2) - \nonumber\\
    &&    N\cdot (A\cdot V_{FS})^2\cdot \frac{(N_S+2)^2}{N_S+N_B}\nonumber\\
&=& (A\cdot V_{FS})^2 \cdot (N_S+2)\cdot \big[\frac{4}{3}\cdot \frac{N_S+1}{N_S}- \frac{N_S+2}{N_S+N_B}\big]\nonumber\\
&=& \frac{4(A\cdot V_{FS})^2(N_S+2)}{3 N_S(N_S+N_B)}\cdot\nonumber\\
&&\big[1/4\cdot N_S^2+N_S N_B+N_B-1/2\cdot N_S \big]
\end{eqnarray}
Then using the leading terms only, one obtains for the position uncertainty:
\begin{eqnarray}
\boxed{
\begin{aligned}
\frac{\sigma_{<x>}}{r}  = \sqrt{\frac{3}{2}}\cdot \big(\frac{\sigma_V}{A\cdot V_{FS}}\big) &\cdot \sqrt{\frac{N_S(N_S+N_B)}{(N_S+2)(\frac{1}{4} N_S^2+N_S N_B )}}\label{EquTriSigma}\\
\textrm{$N_B$=0\,:\quad} \frac{\sigma_{<x>}}{ r} = \sqrt{3} \cdot \sqrt{2}&\cdot \big(\frac{\sigma_V}{A\cdot V_{FS}}\big) \cdot \frac{1}{\sqrt{N_S+2}}\\
\textrm{$N_B$=$N_S$\,:\quad} \frac{\sigma_{<x>}}{ r} =\sqrt{\frac{3}{5}} \cdot 2 &\cdot \big(\frac{\sigma_V}{A\cdot V_{FS}}\big) \cdot \frac{1}{\sqrt{N_S+2}} \label{EquTriSigSB}
\end{aligned}
}
\end{eqnarray}
\\

\subsection{Comparison of Results \& Conclusions}
The main results of the previous sections are compiled for a review of the different approaches. We limit our considerations to the case of the triangular pulse shape and a centred beam position. For off-centre positions the following substitution is required in all equations for position uncertainties:
\begin{equation}
\sqrt{2}\cdot\big(\frac{\sigma_V}{ A \cdot V_{FS}}\big) \rightarrow 4 \cdot\big(\frac{\sigma_V}{V_{FS}}\big) \frac{\sqrt{A_L^2 + A_R^2}}{(A_L+A_R)^2} \nonumber
\end{equation}
{\bf Integral Approach:}\\
\begin{equation}
\frac{\sigma_{<x>}}{ r} =  \sqrt{2} \cdot  \big(\frac{\sigma_V}{A\cdot V_{FS}}\big) \cdot \frac{\sqrt{N_S+N_B}}{ N_S+2}\cdot \sqrt{1+(\frac{N_S+N_B}{N_O})}\nonumber
\end{equation}
The last term describes the influence of the baseline restoration.\\

{\bf RSS Approach or Direct Proportion Fit:}\\
\begin{equation}
\frac{\sigma_{<x>}}{r} =\sqrt{\frac{3}{2}}\cdot \big(\frac{\sigma_{V}}{A \cdot V_{FS}}\big)\cdot\sqrt{\frac{{N_S(N_S+N_B)}}{(N_S+2) \cdot (N_S+1)(\frac{1}{4}N_S+N_B)}}\nonumber
\end{equation}
The last term describes the effect of baseline offset due to AC coupling.\\

{\bf Straight-line Fit:}\\
\begin{equation}
\frac{\sigma_{<x>}}{r} = \sqrt{\frac{3}{2}}\cdot \big(\frac{\sigma_V}{A\cdot V_{FS}}\big) \cdot \sqrt{\frac{N_S(N_S+N_B)}{(N_S+2)\cdot N_S (\frac{1}{4} N_S+N_B )}}\nonumber
\end{equation}
The uncertainty of the straight-line fit is identical to the case of the direct proportion with AC coupling. Note that, in principle, also the uncertainty of the slope parameter can directly be calculated from equation~\ref{EquVarLine} and made available online to the user in order to display the quality of the measurement.\\

Common to all equations is the term $\sigma_V/(A\cdot V_{FS})$. It represents the intrinsic system resolution, i.e. the minimum discernible voltage change at the ADC input, for a given relative signal level $A$. By dividing this factor out, we define the relative uncertainty $(\sigma_{<x>}/r)_{rel}$ which depends only on the analysed sample numbers and is independent of hardware characteristics:\\
\begin{equation*}
(\frac{\sigma_{<x>}}{r})_{rel} = \left(\frac{\sigma_{<x>}}{r}\right)/(\sigma_V/(A\cdot V_{FS}))
\end{equation*}

We had defined $\sigma_V$ as the global sample uncertainty which is composed of the individual noise contributions of the electronics chain. Hence, the noise characteristics of all components, in our case amplifier and ADC, should be properly matched. 

We summarise the most important results:
\begin{itemize}
\item Integral Approach: 
\begin{enumerate} 
\item This approach yields the least stable estimator for beam position and should be abandoned.
\item A clear distinction between baseline samples $N_B$ and signal samples $N_S$ is very important. Baseline samples carry no information and increase the position uncertainty.
\item Further, one needs a large number of offset samples $N_O$ for the baseline restoration. 
\item Any remaining offset in the signal can introduce unknown position biases and also affect derived quantities like the fractional tune.
\item Bunch-by-bunch measurement: A tight window around the signal pulse is required to minimise the position uncertainty.
\end{enumerate}

\item RSS Calculation: 
\begin{enumerate}
\item The RSS calculation yields a stable estimator for the beam position only if no signal offsets exist apart from the offset caused by the AC coupling.
\item No separation between baseline and signal is required. Samples are weighted according to their amplitude.
\item No baseline restoration is required for AC coupled signals. Here, the baseline samples contribute to the position information.
\item Any deviation from the offset due to AC coupling can introduce unknown position biases and also affect derived quantities like the fractional tune.
\item Bunch-by-bunch measurement: The window needs to cover the full signal pulse and may exceed the signal area without adverse effects. 
\item The RSS calculation can be interpreted as a weighted mean of single position measurements.
\item The RSS calculation is equivalent to a fit of a direct proportion.
\end{enumerate}

\item Straight-Line Fit: 
\begin{enumerate}
\item The straight-line fit yields the most stable position estimator. It is immune to offsets in the electrode signals which shift the origin of the coordinate system. This does not change the slope parameter.
\item The position uncertainty can be readily obtained. The only required additional quantity is the variance $\sigma^2_{\Delta}$.
\item Baseline samples add information and improve the knowledge on the "floating" coordinate origin. 
\item The position uncertainty for RSS calculation (direct proportion) and straight-line fit are identical for typical sample numbers $N_S>10$. Only for the case of direct proportion without baseline offset (single-pass BPM in transfer lines), the theoretical value for the position uncertainty is smaller.
\item Bunch-by-bunch measurement: The window needs to cover the full signal pulse and may exceed the signal area. A large window reduces the position uncertainty. 
\item Asynchronous Mode: This new mode seems the "natural" way to analyse BPM data samples. Since the fit does not take reference to any external information (like bunch-based window detection, rf signals, etc.) the beam orbit can be calculated continuously from fixed-size data blocks taken from the incoming data stream of ADC samples. This is somewhat equivalent to listening to the radio.
\item Asynchronous Mode and Bunch-by-bunch measurement: If a gate or window is applied to the continuous data stream of BPM samples, defining a sub-set of the data which is the input to the fitting routine, a bunch-by-bunch measurement is performed.  
\item The asynchronous mode may provide a means to measure the position of a coasting beam, if the noise level can be reduced sufficiently and/or observation window is long enough in order to get access to the Schottky signal.
\item Another advantage of the fit is the independence of the sample order, i.e. the pulse shape, and hence abnormal pulse shapes with irregular structures do not hamper the result. Such cases can happen during mismatched injection into a synchrotron, acceleration or bunch merging. 
\item Gau\ss-Markov Theorem~\cite{9}: Finally, we note that the least-square estimator $<m>$ of the straight-line is a minimum-variance (best), linear and unbiased estimator (BLUE) since the errors $u_i=\Delta-(<m> \Sigma + <c>)$ are uncorrelated and have the same uncertainty $\sigma(i)=\sigma_V$ around the expectation value $E(u_i)=0$.

\end{enumerate}

\end{itemize}

\begin{figure}
\centering
\includegraphics[width=110mm]{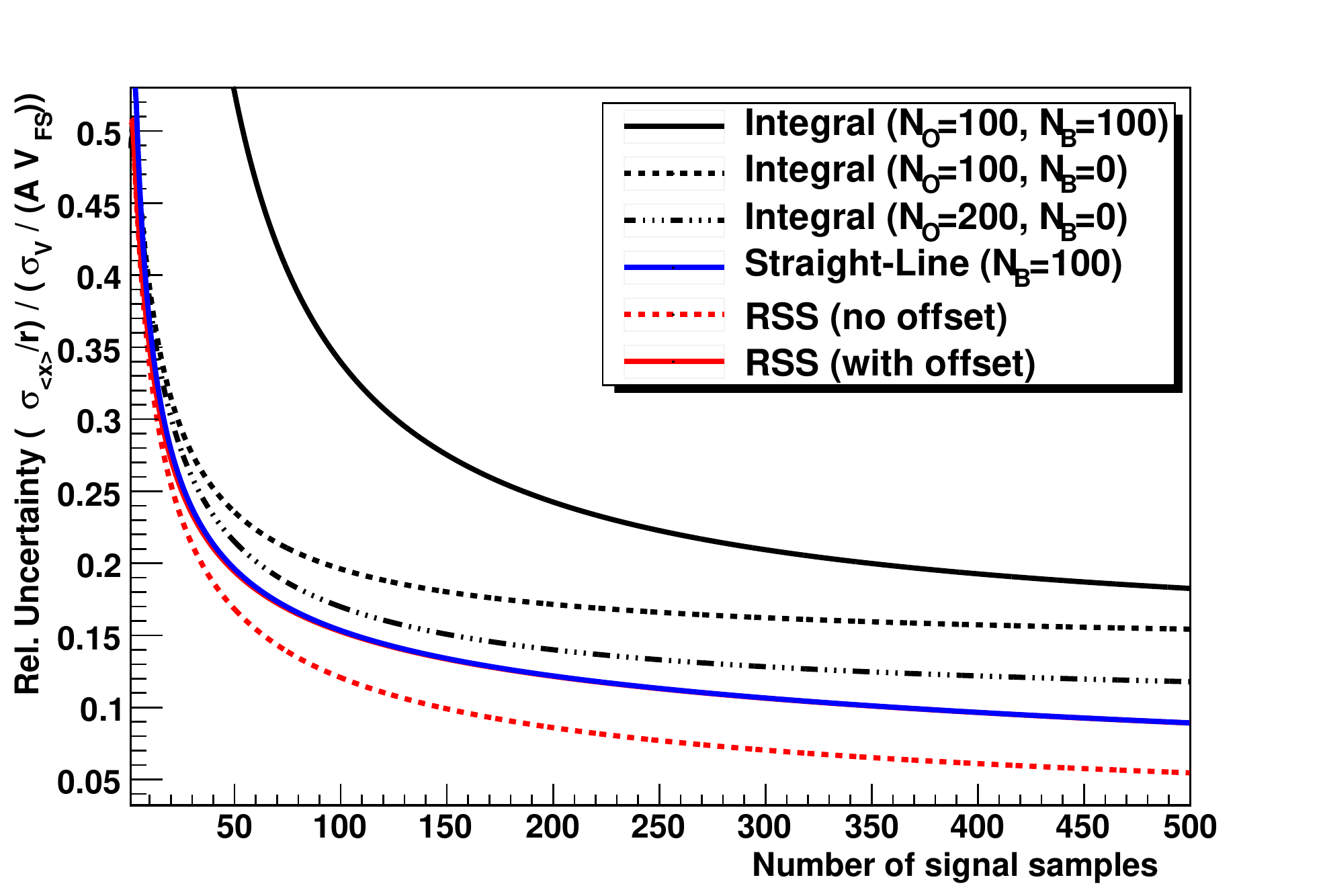}
\caption{Comparison of position uncertainty for all approaches for sample numbers $N_S<500$. The uncertainty is given in relative units in units of $\sigma_V/(A\cdot V_{FS})$.}
\label{fig:UncComp}
\end{figure}

\begin{figure}
\centering
\includegraphics[width=110mm]{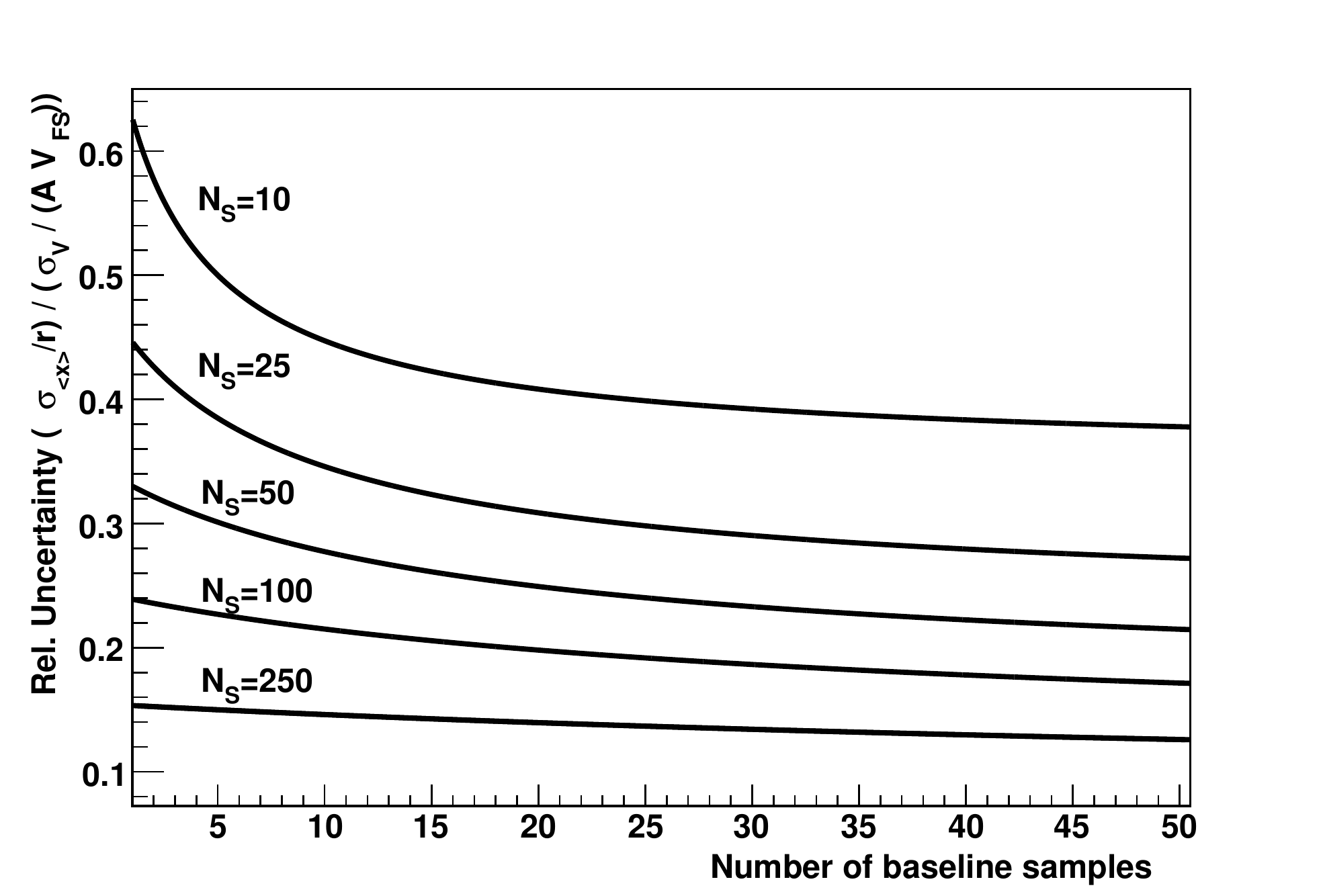}
\caption{Comparison of position uncertainty in the straight-line approach for different sample numbers in the range of 10 to 250 as function of baseline number $N_B$. The uncertainty is given in relative units of $\sigma_V/(A\cdot V_{FS})$.}
\label{fig:UncComp2}
\end{figure}

The uncertainty estimates are compared in Fig.~\ref{fig:UncComp}. RSS (red) and straight-line fit (blue) quickly approach the same value. The RSS uncertainty for the case without baseline offset (dashed red) should be regarded as a theoretical minimum which may be reached in single-pass applications without external signal distortions. The integral approach (black) yields larger position uncertainties. A number of $N_B$=100 and of $N_O=100$ offset samples has been assumed. The dependence on the quality of the baseline restoration and rejection of baseline samples is shown for the combinations of $N_O$=100/$N_B$=0 (dashed) and $N_O=200/N_B=0$ (dot-dashed), respectively. For large sample numbers, the uncertainty contribution due to the baseline restoration dominates the position uncertainty.
In Fig.~\ref{fig:UncComp2} we compare the uncertainties of the straight-line fit for different sample numbers $N_S$ as function of the number of baseline samples $N_B$. The chosen sample numbers in the range of 10 to 250 represent typical bunch lengths and the plot can be used to determine the uncertainty for practical applications. The curves illustrate once more that baseline samples add information to the parameter estimators since the uncertainty is significantly reduced, if $N_B\ge N_S$.

\clearpage
\section{Application to Example Beams}
\label{SecApplication}

It is instructive to calculate the expected position uncertainty for a few given examples in order to quantify the practical requirements on the ADC resolution (and intrinsically the noise limits along entire amplifier chain). To this purpose we apply the results obtained for the fit approach.

\subsection{Position Uncertainty for 50 ns pulse}
Let us look again at a cylindrical BPM with radius $r=50$ mm. We assume the following technical parameters for the ADC: 250 MSa/s sampling speed and bipolar input range $\pm$1 Volt or 2 Volt total input span. During a short 50 ns pulse the ADC acquires $N_S\sim12$ signal samples (we disregard signal distortion in a long transmission line which may "stretch" the signal shape).\\ 
For a conservative or worst-case estimate we set $N_B=N_S$ and to use equation~\ref{EquTriSigSB} of the straight-line fit to calculate the position resolution:
\begin{equation}
\frac{\sigma_{<x>}}{r} =1.55\cdot \big(\frac{\sigma_{V}}{A \cdot V_{FS}}\big)\cdot \frac{1}{\sqrt{N_S+2}}
\end{equation}
For a typical single-pass measurement in a transfer line BPM we set $A=0.5$ and $V_{FS}$=1 Volt, i.e. we consider a unipolar signal since the AC coupling will not produce a significant baseline offset in this case:
\begin{eqnarray}
\frac{\sigma_{<x>}}{r} &=&0.83\cdot \big(\frac{\sigma_{V}}{V_{FS}}\big)\\
\sigma_{<x>}&=& 41 \cdot \big( \frac{\sigma_{V}}{V_{FS}} \big)\,\textrm{mm}
\end{eqnarray}
The common, almost standardised requirement for BPMs is a resolution of 0.1 mm.   One can achieve this value, if the complete measurement system, consisting of amplifier and ADC, provides at least 9 effective bits. For the chosen maximum input level of 1000 mV, this translates to a required sample uncertainty $\sigma_{V}=2.5$ mV.
State-of-the-art hardware stays within those requirements, since 250 MSa/s ADCs feature $\sim$12 effective bits and amplifiers achieve output noise levels of $<$2 mV in a bandwidth of 50 MHz, typical for hadron accelerators.\\
In the case of a ring BPM the AC coupling will result in a bipolar signal and the full ADC input range can be exploited. Hence, one gains one more bit of resolution in the position measurement.

\subsection{BPMs in storage ring CRYRING}
At CRYRING the signal of the 100 mm diameter BPMs are enhanced by a custom-built, low-noise amplifier "CryAmp"~\cite{10}. The CryAmp noise levels can be estimated from the documentation of W. Kaufmann to be:

\begin{itemize}
\item 40 dB; 40 MHz bandwidth: Noise level $\sim$12.5 mV$_{pp}$ or $\sigma_{V}\sim$2.5 mV
\item 60 dB; 40 MHz bandwidth: Noise level $\sim$55 mV$_{pp}$ or $\sigma_{V}\sim$11.0 mV 
\item 60 dB; 4 MHz bandwidth: Noise level $\sim$20 mV$_{pp}$ or  $\sigma_{V}\sim$4.0 mV 
\end{itemize} 

The signals are digitized by 16 bit ADCs of 125 MSa/s sampling frequency with single-ended inputs. The bipolar input range is $\pm$1 Volt.\\
The expected bunch length evolves from $\sim5$ $\mu$s to 150 ns through the acceleration cycle. In this range, the number of acquired samples drops from 625 to 18. Position uncertainties are compiled in table~\ref{tabPosUnc} for a few representative combinations of $N_S$ and $N_B$. For A=0.5 a position resolution well below 0.1 mm can be expected in all cases.

\begin{table}
\caption{Position Uncertainty for different sample configurations.}
\begin{center}
\begin{tabular}{|c|c|c|c|}\hline
   $N_S$ & $N_B$ & $(\sigma_{<x>}/r)_{rel}$ & $\sigma_{<x>}$(40 dB) / mm\\\hline\hline
 18  & 18   &  0.35  &  0.069  \\ \hline
 18  & 282  &  0.28  &  0.056  \\ \hline
 300 & 300  &  0.090 &  0.018  \\ \hline
 625 & 0    &  0.098 &  0.020  \\ \hline
 625 & 75   &  0.085 &  0.017  \\ \hline
\end{tabular}\label{tabPosUnc}
\end{center}
\end{table}


\subsection{Model Comparison with Simulated Beam}

\begin{figure}
\centering
\includegraphics[width=115mm]{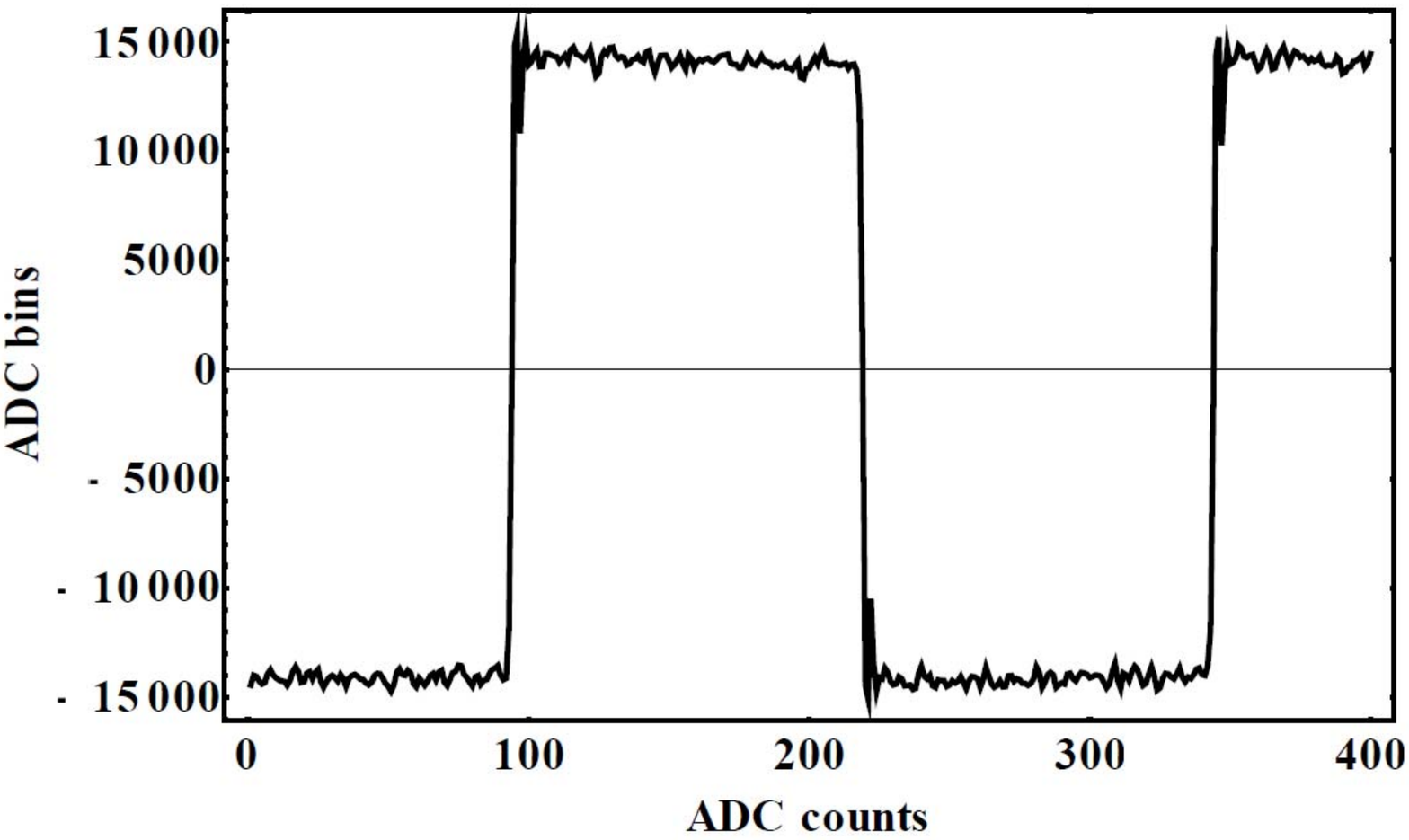}
\caption{Example of square input signal fed to ADC system of 16 nominal bits. Noise level $\sigma_V\approx$ 30 mV.}
\label{fig:UncCompRssInput}
\end{figure}

We present a comparison between theoretical model and real ADC data, acquired by a 250 MSa/s, 12 eff. bit ADC system, for two cases of 6 and 17 mm offset. Two symmetric square signals of 50\% duty ($N_S$=125, $N_B$=125) were generated by an arbitrary function generator as shown in figure~\ref{fig:UncCompRssInput}. White noise of 30 mV(rms) was added independently to both signals. Around the edges some ringing is visible. 

\begin{figure}
\centering
\includegraphics[width=115mm]{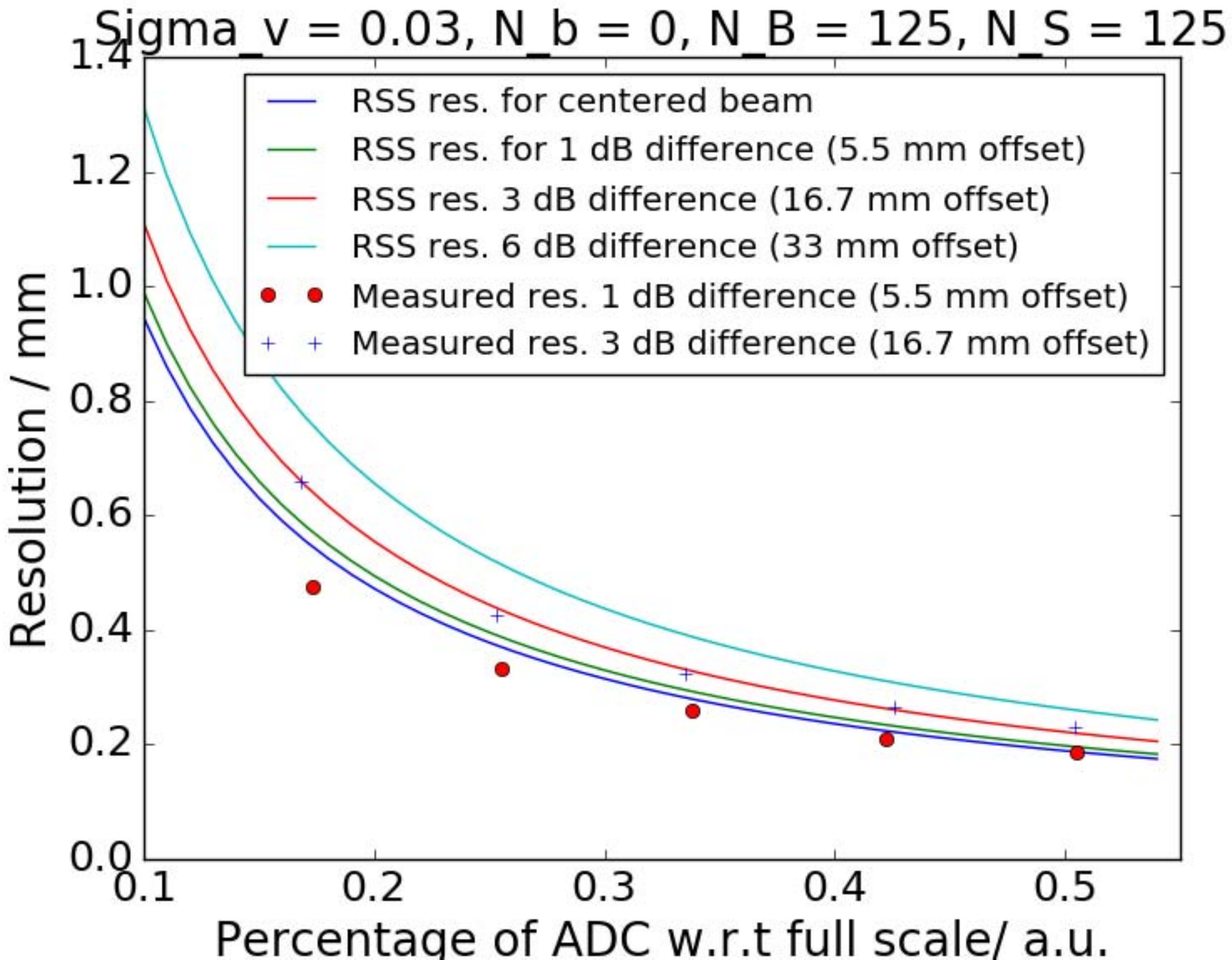}
\caption{Comparison of position uncertainty; baseline excluded. Data taken by a Libera ADC system are compared to the model calculation (equ.~\ref{EquUncRssSquGen}).  For explanation see text.}
\label{fig:UncCompRssLibera2}
\end{figure}

\begin{figure}
\centering
\includegraphics[width=115mm]{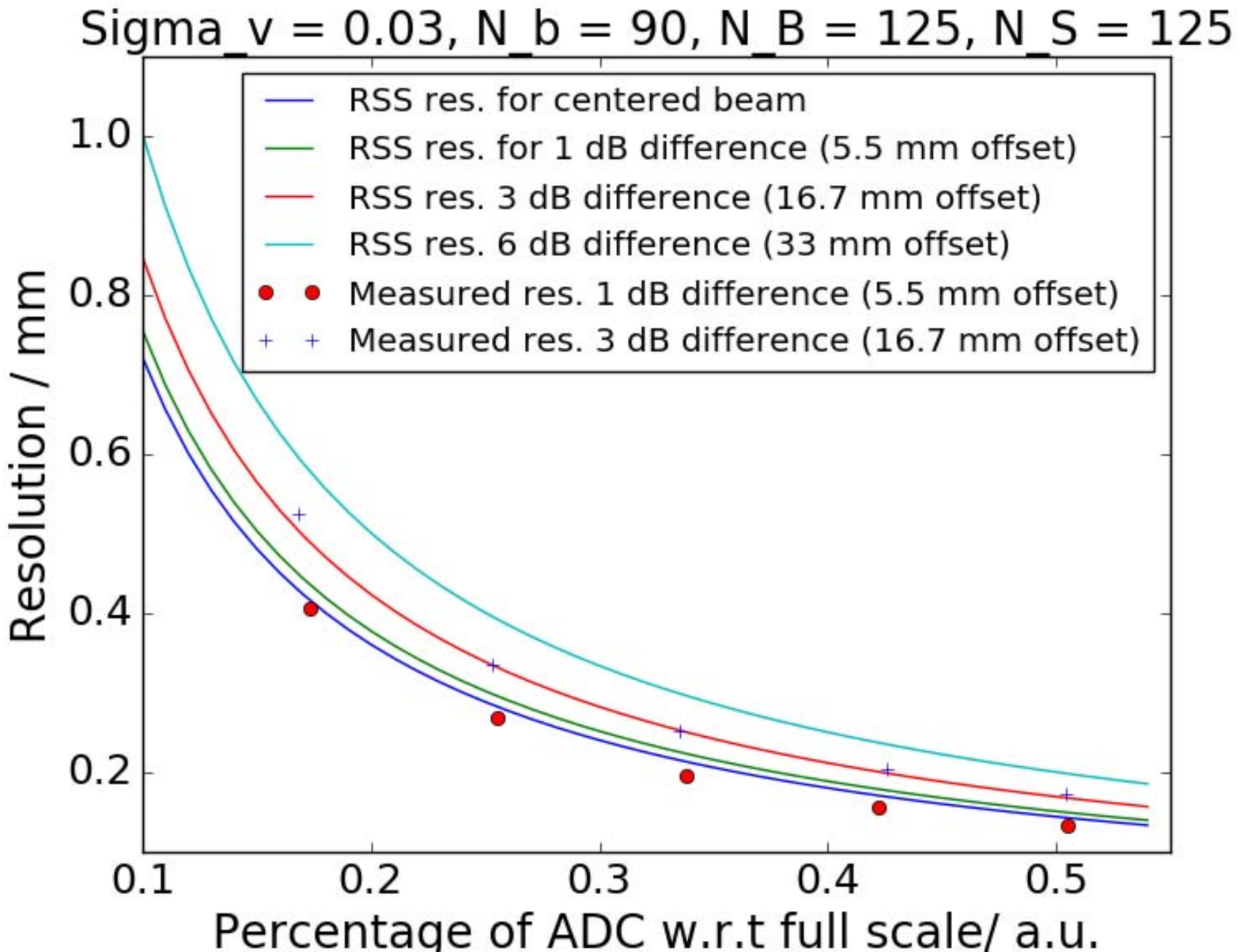}
\caption{Comparison of position uncertainty; most of baseline included. Data taken by a Libera ADC system are compared to the model calculation (equ.~\ref{EquUncRssSquGen}). For explanation see text.}
\label{fig:UncCompRssLibera}
\end{figure}

Position offsets were simulated by attenuating one of the signals with respect to the reference signal whose signal amplitude is plotted in figures~\ref{fig:UncCompRssLibera2} and~\ref{fig:UncCompRssLibera}. In the former case only the signal part (positive values) have been analysed, while in the latter case also $N_b$=90 baseline samples have been added. For 6 mm offset and excluded baseline the data initially fall below the model prediction for A $<$ 0.2 (about 10-15\%) and then approach the model for larger signal levels. When the baseline is included, the uncertainty is significantly reduced. The data fall only slightly below the model prediction at all amplitudes. For the 17 mm offset there is very good agreement for all signal levels in both analysis cases.

\clearpage

\section{Appendix}
\subsection{Summary of Equations}

\begin{sidewaysfigure}
\begin{tabular}{| l | c |}

    \hline
    Square pulse without baseline offset (single pulse)& \\
    Amplitudes refer to black coordinate system & \\[-0.5cm]
     \includegraphics[height=50 mm]{SquarePulseAC}  & \begin{minipage}[b]{.5\textwidth}
  Integral: All samples have equal weights
  \begin{eqnarray}
\frac{\sigma_{<x>}}{ r} &=&   2 \cdot  \frac{\sigma_V}{ V_{FS}} \cdot \frac{\sqrt{A_L^2 + A_R^2}}{(A_L + A_R)^2} \cdot \frac{\sqrt{N_S+N_B}}{ N_S} \nonumber
\end{eqnarray}  
RSS: Baseline samples do not contribute (weight = 0)
\begin{eqnarray}
\frac{\sigma_{<x>}}{r} &=& 2 \cdot \frac{\sigma_V}{V_{FS}} \cdot \frac{\sqrt{A_L^2 + A_R^2}}{(A_L+A_R)^2} \cdot \frac{1}{\sqrt{N_S} } \nonumber
\end{eqnarray}
Fit: 
\begin{eqnarray}
\frac{\sigma_{<x>}}{r}   &=& 2 \cdot \frac{\sigma_V}{V_{FS}} \cdot \frac{\sqrt{A_L^2 + A_R^2}}{(A_L+A_R)^2}  \cdot \frac{\sqrt{N_S+N_B}}{\sqrt{N_S \cdot N_B}} \nonumber
\end{eqnarray}

\end{minipage}  \\ \hline
      Triangular pulse without baseline offset (single pulse)& \\
      Amplitudes refer to black coordinate system & \\[-0.5cm]
     \includegraphics[height=50 mm]{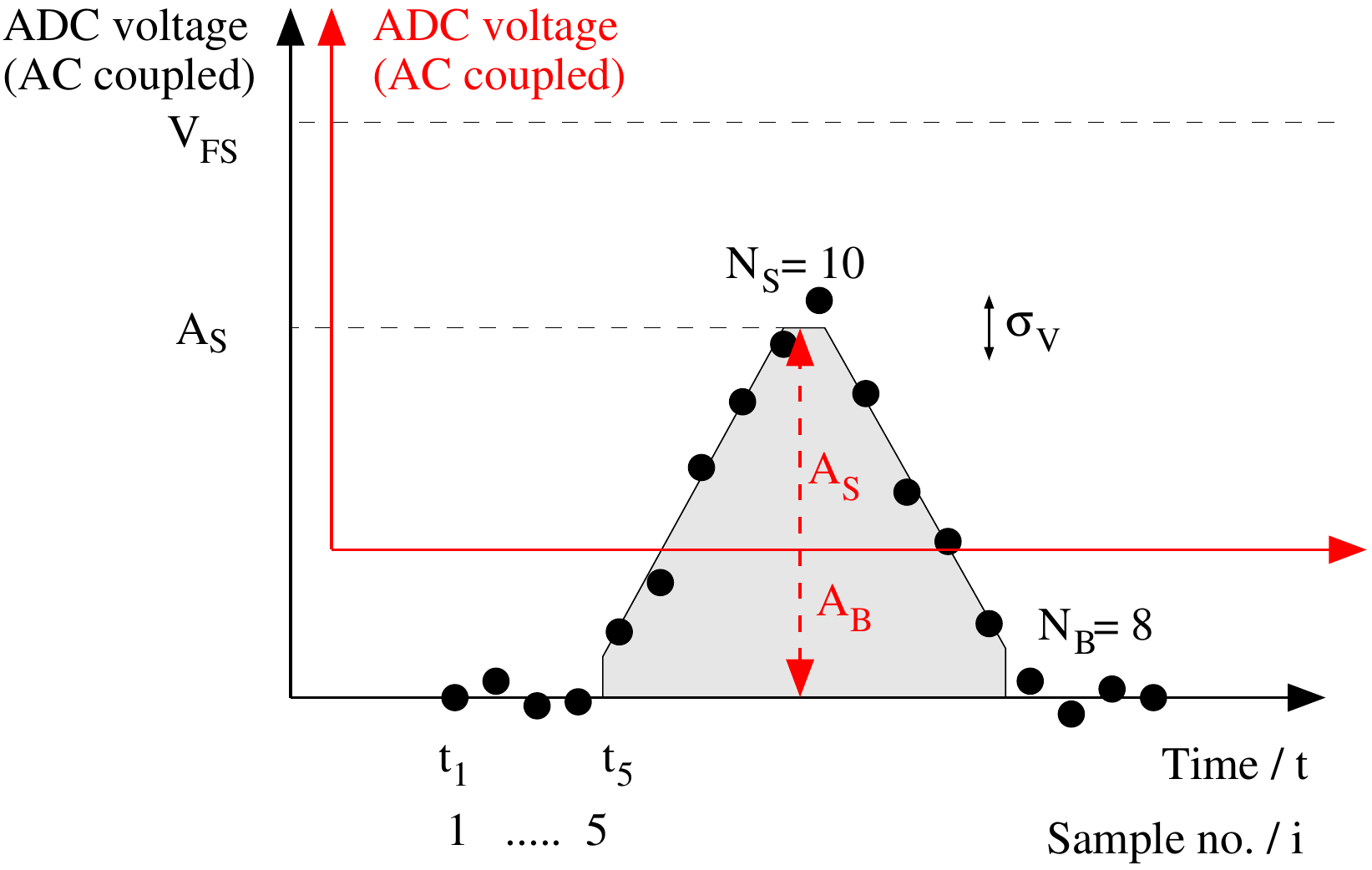} &\begin{minipage}[b]{.5\textwidth}%
     Integral: 
     \begin{eqnarray}
\frac{\sigma_{<x>}}{ r} &=&   4 \cdot  \frac{\sigma_V}{ V_{FS}} \cdot \frac{\sqrt{A_L^2 + A_R^2}}{(A_L + A_R)^2} \cdot \frac{\sqrt{N_S+N_B}}{ N_S+2} \nonumber
\end{eqnarray}
RSS: 
\begin{equation}
\frac{\sigma_{<x>}}{r} = 2 \sqrt{3} \cdot  \frac{\sigma_V}{ V_{FS}} \cdot \frac{\sqrt{A_L^2 + A_R^2}}{(A_L + A_R)^2}  \cdot \sqrt{\frac{N_S}{(N_S+2)(N_S+1)}} \nonumber
\end{equation}
Fit:
\begin{eqnarray}
\frac{\sigma_{<x>}}{r}  = 2 \sqrt{3}  \cdot  \frac{\sigma_V}{ V_{FS}} \cdot \frac{\sqrt{A_L^2 + A_R^2}}{(A_L + A_R)^2}  \cdot \sqrt{\frac{N_S(N_S+N_B)}{(N_S+2)(\frac{1}{4} N_S^2+N_S N_B )}} \nonumber
\end{eqnarray}
     \end{minipage}  \\ \hline

\end{tabular}
\label{fig:Summary1}
\end{sidewaysfigure}


\begin{sidewaysfigure}

\begin{tabular}{| l | c |}

    \hline
    
    Square pulse with baseline offset due to cyclic pulses& \\
    Turn-by-turn analysis (h=1): &\\
    Analyse all signals in one period $N_S$ and $N_B$ &\\
    Amplitudes refer to black coordinate system without offset & \\[-1.1cm]
     \includegraphics[height=50 mm]{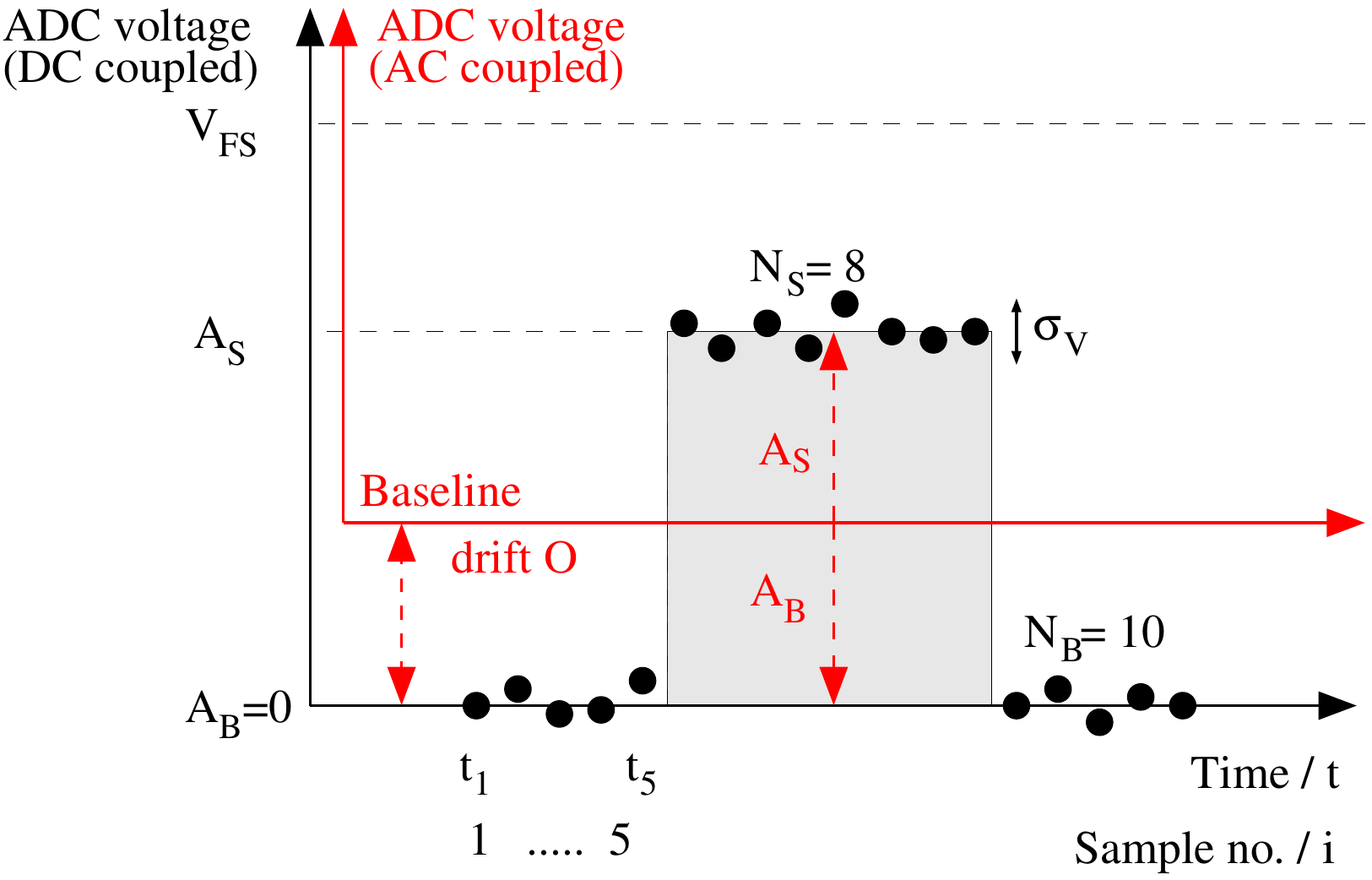}  & \begin{minipage}[b]{.5\textwidth}
  Integral: $N_O$ samples used for offset calculation 
  \begin{eqnarray}
\frac{\sigma_{<x>}}{r} =   \frac{2\cdot \sigma_V}{ V_{FS}} \frac{\sqrt{A_L^2+A_R^2}}{(A_L + A_R)^2} \cdot \frac{1}{N_S} \sqrt{(N_S+N_B) + \frac{(N_S+N_B)^2}{N_O}} \nonumber
\end{eqnarray}  
RSS: Balanced system - $N_B$ samples between pulses (fit through origin)
\begin{eqnarray}
\frac{\sigma_{<x>}}{r} = \frac{ 2 \cdot\sigma_V}{V_{FS}}  \frac{\sqrt{A_L^2 + A_R^2}}{(A_L+A_R)^2} \cdot \sqrt{\frac{N_S+N_B}{N_S \cdot N_B} } \nonumber
\end{eqnarray}
Fit: 
\begin{eqnarray}
\frac{\sigma_{<x>}}{r}   =  \frac{ 2 \cdot\sigma_V}{V_{FS}} \frac{\sqrt{A_L^2 + A_R^2}}{(A_L+A_R)^2}  \cdot \sqrt{\frac{N_S+N_B}{N_S \cdot N_B}} \nonumber
\end{eqnarray}

\end{minipage}  \\ \hline
      Triangular pulse with baseline offset due to cyclic pulses & \\
      Turn-by-turn analysis (h=1): &\\
      Analyse all signals in one period $N_S$ and $N_B$ &\\
      Amplitudes refer to black coordinate system without offset & \\[-1.1cm]
     \includegraphics[height=50 mm]{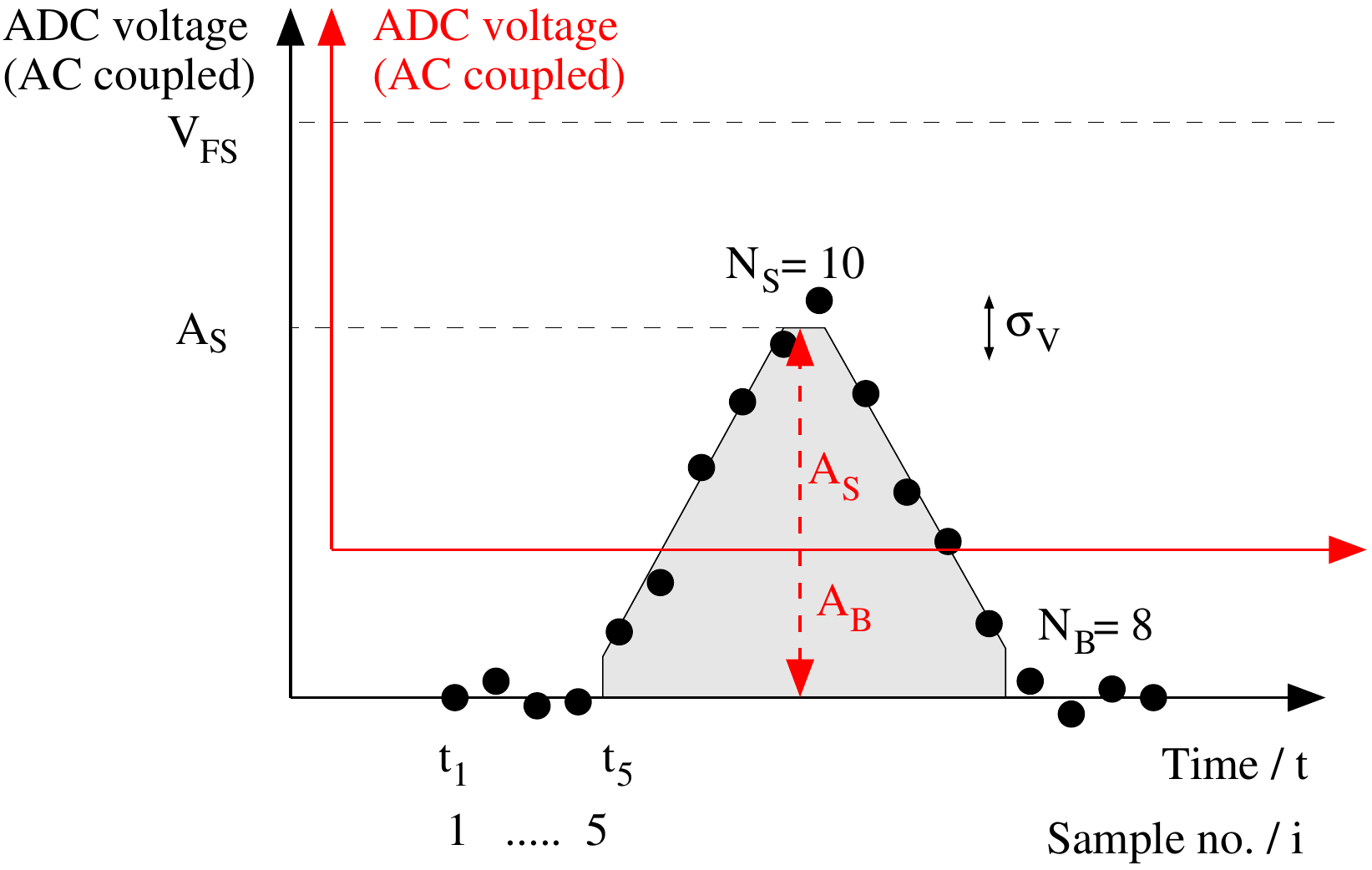} &\begin{minipage}[b]{.5\textwidth}%
     Integral: 
     \begin{eqnarray}
\frac{\sigma_{<x>}}{ r} =   \frac{ 4 \cdot \sigma_V}{ V_{FS}} \frac{\sqrt{A_L^2 + A_R^2}}{(A_L + A_R)^2}\cdot \frac{1}{N_S} \sqrt{(N_S+N_B) + \frac{(N_S+N_B)^2}{N_O}}  \nonumber
\end{eqnarray}
RSS: Balanced system - $N_B$ samples between pulses 
\begin{eqnarray}
\frac{\sigma_{<x>}}{r} &\approx&  \frac{  2 \sqrt{3}  \cdot\sigma_V}{ V_{FS}} \cdot \frac{\sqrt{A_L^2 + A_R^2}}{(A_L + A_R)^2} \nonumber\\
&\cdot& \sqrt{\frac{N_S(N_S+N_B)}{(N_S+2)(\frac{1}{4}N_S^2+N_SN_B + N_B + \frac{1}{4}N_S)}} \nonumber
\end{eqnarray}
Fit: Identical for RSS for leading terms of $N_S$ and $N_B$
\begin{eqnarray}
\frac{\sigma_{<x>}}{r}  \approx    \frac{ 2 \sqrt{3}  \cdot\sigma_V}{ V_{FS}} \cdot \frac{\sqrt{A_L^2 + A_R^2}}{(A_L + A_R)^2}  \cdot \sqrt{\frac{N_S(N_S+N_B)}{(N_S+2)(\frac{1}{4} N_S^2+N_S N_B )}} \nonumber
\end{eqnarray}
     \end{minipage}  \\ \hline

\end{tabular}
\label{fig:Summary2}
\end{sidewaysfigure}


\begin{sidewaysfigure}

\begin{tabular}{| l | c |}

    \hline
    
    Square pulse with baseline offset & \\
    Gated analysis: symmetric window around signal pulse & \\ 
    Amplitudes refer to black coordinate system without offset & \\
    \includegraphics[height=50 mm]{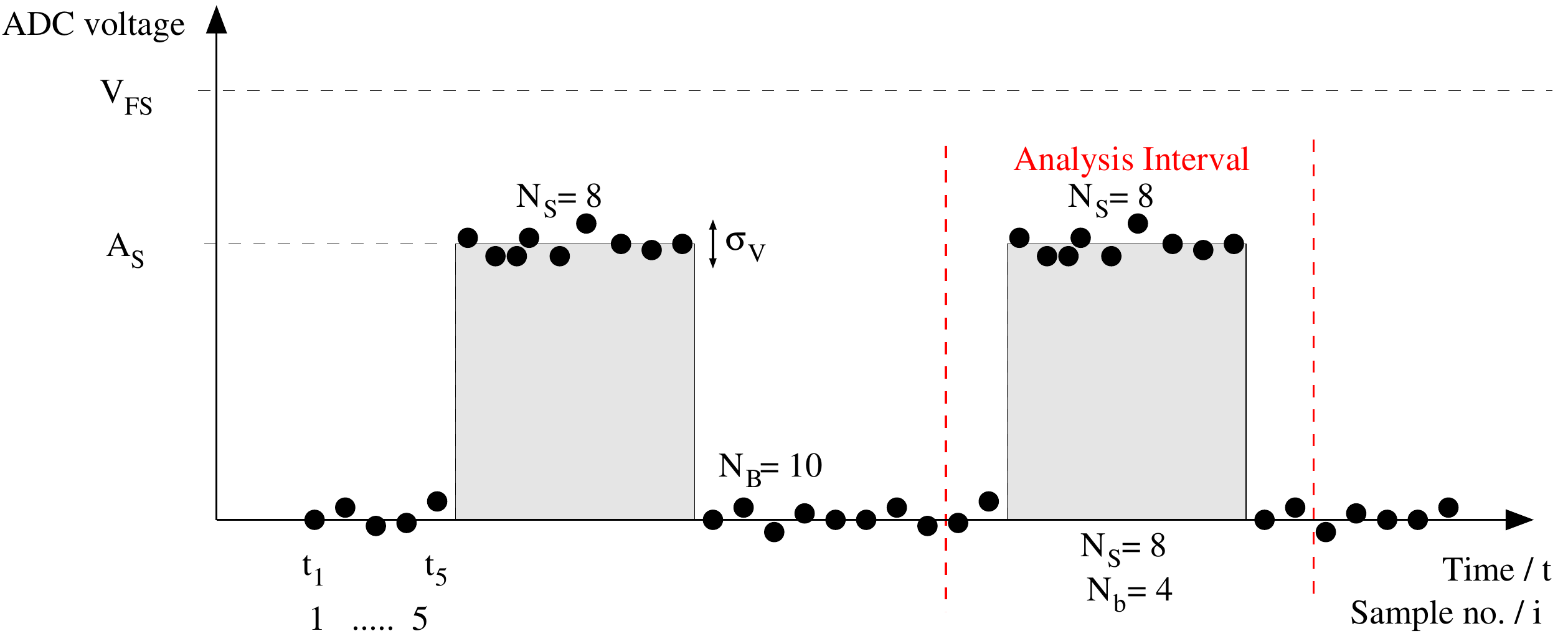} & \begin{minipage}[b]{.5\textwidth}
  Integral: see case with offset, $N_b$=$N_B$\\[1cm]
RSS: Balanced system - $N_B$ samples between pulses, but smaller analysis window 
\begin{eqnarray}
\frac{\sigma_{<x>}}{r} &=& 2 \cdot \frac{\sigma_V}{V_{FS}} \cdot \frac{\sqrt{A_L^2 + A_R^2}}{(A_L+A_R)^2} \cdot \frac{(N_S+N_B)}{\sqrt{N_S(N_b \cdot N_S+N_B^2)} }\nonumber
\end{eqnarray}
Fit: 
\begin{eqnarray}
\frac{\sigma_{<x>}}{r}   =  \frac{ 2 \cdot\sigma_V}{V_{FS}} \frac{\sqrt{A_L^2 + A_R^2}}{(A_L+A_R)^2}  \cdot \frac{\sqrt{N_S+N_b}}{\sqrt{N_S \cdot N_b}} \nonumber
\end{eqnarray}

\end{minipage}  \\ \hline
      Triangular pulse with baseline offset & \\
      Gated analysis: symmetric window around signal pulse & \\ 
       Amplitudes refer to black coordinate system without offset & \\
      \includegraphics[height=50 mm]{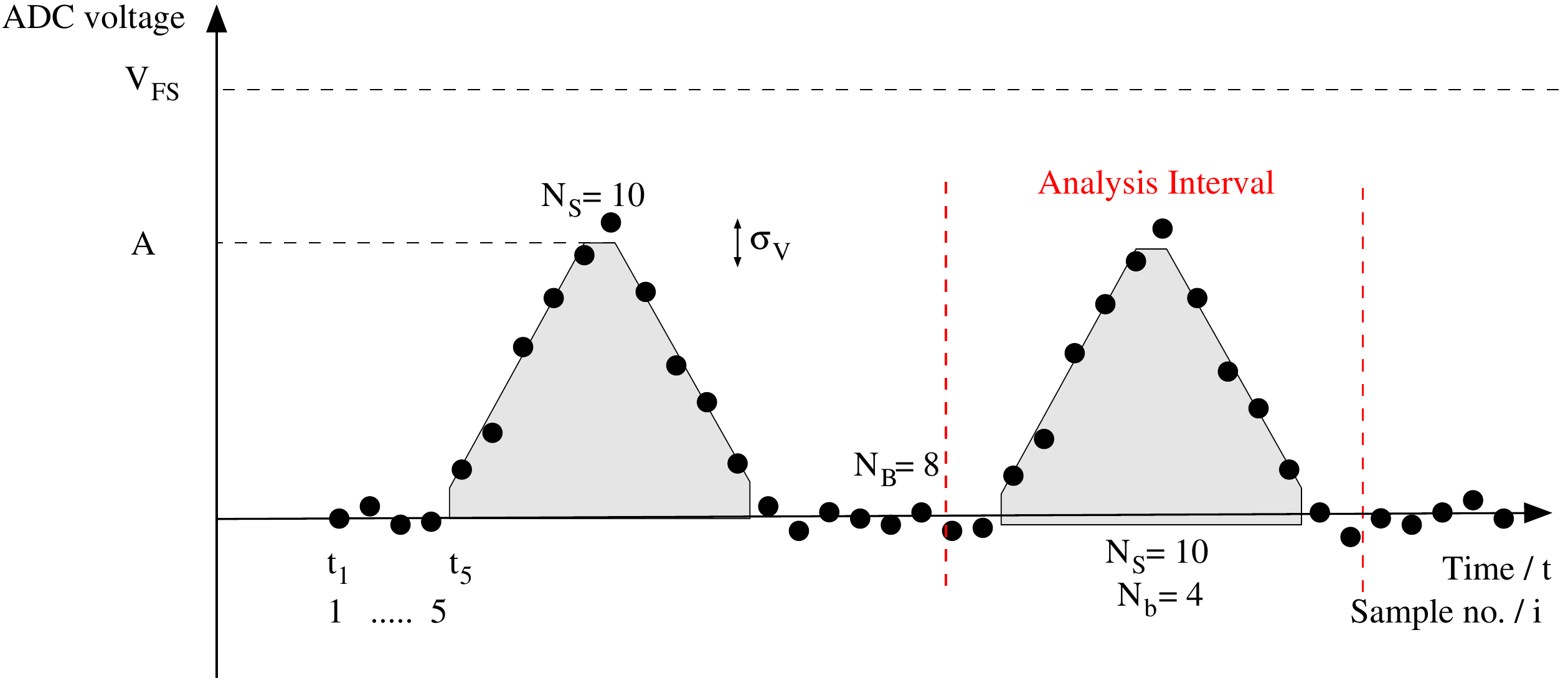} &\begin{minipage}[b]{.5\textwidth}%
     Integral: see case with offset, $N_b$=$N_B$\\[1cm]
RSS: Balanced system - $N_B$ samples between pulses
\begin{eqnarray}
\frac{\sigma_{<x>}}{r} &\approx&  \frac{  2 \sqrt{3}  \cdot\sigma_V}{ V_{FS}} \cdot \frac{\sqrt{A_L^2 + A_R^2}}{(A_L + A_R)^2} \cdot \sqrt{\frac{N_S}{(N_S+2)(N_S+1)}}\nonumber
\\ &&\cdot \frac{1}{\sqrt{
1-\frac{3}{4} \frac{N_S}{N_S+N_B}\cdot (1 - \frac{N_b-N_B}{N_S+N_B})}} \nonumber
\end{eqnarray}
Fit:
\begin{eqnarray}
\frac{\sigma_{<x>}}{r}  \approx    \frac{ 2 \sqrt{3}  \cdot\sigma_V}{ V_{FS}} \cdot \frac{\sqrt{A_L^2 + A_R^2}}{(A_L + A_R)^2}  \cdot \sqrt{\frac{N_S(N_S+N_b)}{(N_S+2)(\frac{1}{4} N_S^2+N_S N_b }} \nonumber
\end{eqnarray}
     \end{minipage}  \\ \hline

\end{tabular}
\label{fig:Summary3}
\end{sidewaysfigure}

\newpage
\subsection{List of Symbols}
\begin{table}[h!]
\begin{center}
\begin{tabular}{|l|l|l|}\hline
Variable & Symbol & Comment \\ \hline
Beam position & $x$ & \\\hline
Estimator of beam position & $<x>$ & \\\hline
Mean beam position & $\overline{x}$  & \\\hline
Number of signal samples & $N_S$ &  \\\hline
Number of baseline samples & $N_B$ &  \\\hline
Number of analysed baseline samples & $N_b$ & \\ \hline
Sample time & $t_{Sa}$ & \\\hline
Sample number index & i & \\\hline
Full scale voltage & $V_{FS}$& \\\hline
Left electrode signal trace & $S_L$ & \\\hline
Right electrode signal trace & $S_R$ & \\\hline
Offset level in signal & $O$ &  \\\hline 
Integral of left electrode signal & $I_L$ & \\\hline
Integral of right electrode signal & $I_R$ & \\\hline
Root-sum-square left signal & $RSS_L$ & \\\hline
Root-sum-square right signal & $RSS_R$ & \\\hline
Maximum amplitude left signal / $V_{FS}$& $A_{L} \in [0,1]$ & \\\hline
Maximum amplitude right signal / $V_{FS}$& $A_{R} \in [0,1]$ & \\\hline
Difference signal  & $\Delta = S_L - S_R $ & \\\hline
Sum signal & $\Sigma= S_L + S_R $ &   \\\hline
Full scale voltage & $V_{FS}$  & \\\hline
RMS noise voltage of ADC sample & $\sigma_V$ & \\\hline
RMS noise voltage of variable $\Delta$ & $\sigma_\Delta$ & \\\hline
RMS noise voltage of variable $\Sigma$ & $\sigma_\Sigma$ & \\\hline

\end{tabular}
\end{center}
\end{table}


\end{document}